\documentclass[final,5p,times,twocolumn]{elsarticle}
\usepackage{changes}
\usepackage{amsmath,amssymb,bm}
\usepackage{booktabs}
\usepackage{multirow}
\usepackage{physics}
\usepackage[hidelinks]{hyperref}
\journal{Journal of High Energy Astrophysics}
\biboptions{sort&compress}

\providecommand{\varpsi}{\psi}

\RequirePackage{color}\definecolor{RED}{rgb}{1,0,0}\definecolor{BLUE}{rgb}{0,0,1} 
\DeclareOldFontCommand{\sf}{\normalfont\sffamily}{\mathsf} 
\providecommand{\DIFaddtex}[1]{{\protect\color{red} #1}} 
\providecommand{\DIFaddbegin}{} 
\providecommand{\DIFaddend}{} 
\providecommand{\DIFaddFL}[1]{\DIFadd{#1}} 
\providecommand{\DIFaddbeginFL}{} 
\providecommand{\DIFaddendFL}{} 
\providecommand{\DIFadd}[1]{\texorpdfstring{\DIFaddtex{#1}}{#1}} 
\RequirePackage{listings} 
\RequirePackage{color} 
\lstdefinelanguage{DIFcode}{ 
  moredelim=[il][\color{red}\scriptsize]{\%DIF\ <\ }, 
  moredelim=[il][\color{red}]{\%DIF\ >\ } 
} 
\lstdefinestyle{DIFverbatimstyle}{ 
	language=DIFcode, 
	basicstyle=\ttfamily, 
	columns=fullflexible, 
	keepspaces=true 
} 
\lstnewenvironment{DIFverbatim}{\lstset{style=DIFverbatimstyle}}{} 
\lstnewenvironment{DIFverbatim*}{\lstset{style=DIFverbatimstyle,showspaces=true}}{} 
\begin{document}

\begin{frontmatter}

\title{Energy Loss of Newborn Magnetars by Schwinger Process}

\author[ibs,gistapri,gistdpps]{Chul Min Kim}
\ead{chulmin@gist.ac.kr}

\author[ibs,icranet,kunsan]{Sang Pyo Kim}
\ead{sangkim@kunsan.ac.kr}

\author[icranet,icra,inafrome]{Remo Ruffini}
\ead{ruffini@icra.it}

\author[icranet,icra,inafteramo]{Yu Wang}
\ead{yu.wang@icranet.org}

\author[icranet]{Shurui Zhang\corref{cor1}}
\ead{zhangsr@mail.ustc.edu.cn}
\cortext[cor1]{Corresponding author.}

\address[ibs]{Center for Relativistic Laser Science, Institute for Basic Science, Gwangju 61005, South Korea}
\address[gistapri]{Advanced Photonics Research Institute, Gwangju Institute of Science and Technology, Gwangju 61005, South Korea}
\address[gistdpps]{Department of Physics and Photon Science, Gwangju Institute of Science and Technology, Gwangju 61005, South Korea}
\address[icranet]{ICRANet, 65122 Piazza della Repubblica, 10, Pescara, Italy}
\address[kunsan]{School of Advanced Science and Technology, Kunsan National University, Kunsan 54150, South Korea}
\address[icra]{ICRA, Dip. di Fisica, Sapienza Universit\`a\ di Roma, Piazzale Aldo Moro 5, I-00185 Roma, Italy}
\address[inafrome]{INAF, Viale del Parco Mellini 84, 00136 Rome, Italy}
\address[inafteramo]{INAF -- Osservatorio Astronomico d'Abruzzo, Via M. Maggini snc, I-64100, Teramo, Italy}

\begin{abstract}
We investigate electron--positron pair creation through the Schwinger process in newborn magnetars with millisecond spin periods and surface dipole fields close to or above the QED critical field, $B_{\rm Q} = 4.414\times10^{13}\,\mathrm{G}$. We use the aligned vacuum-rotator field as an unscreened reference and derive the analytical global pair creation flux and recast it into a compact form with accurate analytic approximations. For a fiducial model with $B_{\rm p} = 10^{14}\,\mathrm{G}$ and $P_0 = 1\,\mathrm{ms}$, the initial unscreened rate exceeds the classical Goldreich--Julian open-field particle flux by many orders of magnitude. We treat this rate as an upper bound on the first discharge, not as a long-lived magnetospheric state. The active height of the discharge is estimated from the logarithmic scale height of the polar Schwinger rate, giving $\ell_{\rm Sch}\simeq4.13\times10^4\,\mathrm{cm}$ for the fiducial model. In this region, the charge density needed to screen $E_{\parallel}$ is supplied on $t_{\rm fill}\sim6\times10^{-32}\,$s, while the pair-plasma response time is $\omega_{\rm p}^{-1}\sim4\times10^{-14}\,$s; both are far shorter than the light-cylinder crossing time. We therefore introduce a screening duty factor $f_{\rm scr}=t_{\rm scr}/t_{\rm rep}$ and use it to compute an effective Schwinger luminosity. For the fiducial global-circuit prescription, $t_{\rm rep}=R_{\rm LC}/c=\Omega^{-1}$, $f_{\rm scr}\simeq2.5\times10^{-10}$ at birth, reducing the sustained Schwinger luminosity from $3.5\times10^{58}$ to $8.6\times10^{48}\,\mathrm{erg\,s^{-1}}$. With this factor included, the spin evolution is still affected by the initial discharge but no longer assumes a persistent vacuum magnetosphere. We numerically investigate the Schwinger process in misaligned magnetars and find a multiple structure for large inclination angles.  The maxima of pair production decrease up to three order as the inclination angle increases from $0^\circ$ to $90^\circ$, and their peaks are located between the rotation and magnetic axes. 
\end{abstract}


\end{frontmatter}

\section{Introduction}

Magnetars are neutron stars with ultrastrong magnetic fields, with inferred surface dipole strengths of $10^{14}$--$10^{15}\,$G. Newborn proto magnetars may reach even larger internal and surface fields for a short time through differential rotation and dynamo amplification \citep{1993ApJ...408..194T,2006RPPh...69.2631H,2014ApJS..212....6O,2017ARA&A..55..261K}. Millisecond magnetars have long been discussed as central engines of energetic transients, including gamma ray bursts (GRBs) and relativistic outflows \citep{1992Natur.357..472U,2007ApJ...659..561M,2015PhR...561....1K,2024ApJ...974...89W,2025ApJ...986...14H}. Recent detections of TeV emission from GRBs have renewed interest in the earliest stages of pair loading and energy extraction in these systems \citep{2019Natur.575..455A,2019Natur.575..459A,2019Natur.575..464A,2021Sci...372.1081A,2023Sci...380.1390L}. Recently, a newborn magnetar central engine with $P=4.2\pm0.2\,$ms and $B=(1.6\pm0.1)\times10^{14}\,$G was inferred for a superluminous supernova, suggesting that the millisecond, supercritical regime considered here may indeed occur in nature \citep{2026Natur.651..321F}.

In the standard pulsar and magnetar picture, the magnetosphere is supplied by the Goldreich--Julian (GJ) corotation density, charge extraction from the surface, and pair cascades driven by gap acceleration and high energy photon conversion \citep{1969ApJ...157..869G,1975ApJ...196...51R,1982ApJ...252..337D,1983ApJ...273..761D,2001ApJ...547..929B,2013MNRAS.429...20T}. In that framework, plasma screens large parallel electric fields and supports the current system. Newborn magnetars may also approach a different threshold. If rapid rotation and strong magnetization generate an electric field with a component parallel to the magnetic field, the system can enter the strong field quantum electrodynamics (QED) regime described by Heisenberg, Euler, and Schwinger \citep{1936ZPhy...98..714H,1951PhRv...82..664S,2010PhR...487....1R}. In that regime, charges need not come only from surface extraction; they can be produced directly from the vacuum.

This paper addresses two questions. First, what is the initial Schwinger pair-creation rate in an aligned vacuum dipole before plasma screening, and how does this upper-bound estimate compare with the classical Goldreich--Julian flux? Second, when local screening is represented by a replenishment duty factor, what effective Schwinger torque is obtained compared with magnetic dipole radiation and gravitational wave emission from a deformed newborn magnetar?

The paper is organized as follows. Section~\ref{sec:model} summarizes the aligned vacuum dipole field, the local Schwinger rate, and the global pair flux. 
Section~\ref{sec:screening} gives the screening-time estimate, defines the replenishment duty factor, and sets the range of applicability of the vacuum rate. Section~\ref{sec:gj} compares the resulting near-surface pair-production capacity with the Goldreich--Julian flux. Section~\ref{sec:misaligned} examines the corresponding unscreened rate for misaligned rotators. Section~\ref{sec:spindown} adds the screened Schwinger channel to magnetic dipole and gravitational wave spin evolution and presents both birth parameter maps and a fiducial screened evolution in time. Section~\ref{sec:implications} discusses caveats and astrophysical implications. Section~\ref{sec:conclusions} summarizes the main results.

\section{Schwinger Process from Vacuum Dipole Model for New-born Magnetars}\label{sec:model}
Recently, it has been proposed that a newborn magnetar may have a millisecond period and a magnetic field above the Schwinger (critical) field for electron, $B_{\rm Q} = m_{\rm e}^2 c^3/(e \hbar) = 4.414 \times 10^{13}$ G~\citep{farah2026lense}. In the Schwinger field, the lowest Landau level energy of an electron equals the rest mass energy, and QED effects become important. One of the most significant phenomena is the pair creation of electrons and positrons provided that a background electric field is near or above the Schwinger field $E_{\rm Q} = m_{\rm e}^2 c^3/(e \hbar)$ (for a review, see~\citep{2010PhR...487....1R} ). Vacuum birefringence is another important QED phenomenon to be measured by magnetar X-ray polarometry~\citep{heyl2002qed,taverna2022polarized,2023EPJC...83..104K}.
The Schwinger process from magnetars was studied in~\citep{lieu2017fast}, and, in particular, millisecond magnetars have been shown to lead to catastrophic emission of electrons and positrons~\citep{2023ARep...67S.122K}.

To find the initial unscreened emission of electrons and positrons from millisecond newborn magnetars, we will employ the aligned vacuum dipole (rotator) model introduced by Goldreich and Julian~\citep{1969ApJ...157..869G}. There, an electric field induced by a rotating magnetic dipole has a component parallel to the magnetic field, whose product is strong enough to produce electron-positron pairs near the north and south poles~\citep{kim2006schwinger}; for a recent related application of a rotating magnetic-dipole configuration and its dyadoregion in a compact-merger context, see Appendix~A of Ref.~\cite{2026JHEAp..5000464R}. The electric field used below is the exterior vacuum electrostatic solution for a perfectly conducting rotating star. It is obtained by solving Laplace's equation outside the star, with the potential set to zero at infinity and with the tangential surface field fixed by the corotation boundary condition, ${\bf E}_{t}= -[(\boldsymbol{\Omega}\times{\bf r})\times{\bf B}]_{t}/c$ at $r=R$. It should not be confused with the local ideal-corotation expression extended through the exterior magnetosphere.
The physics of strong field QED action is described by Heisenberg-Euler and Schwinger, which is an exact one-loop effective action in a constant electromagnetic field, which depends only on the Maxwell scalar ${\cal F} = F_{\mu \nu} F^{\mu \nu}/4=\left(\mathbf{B}^2-\mathbf{E}^2\right)/2$ and pseudo-scalar 
${\cal G} = F_{\mu \nu} F^{*\mu \nu}/4=-\mathbf{E}\cdot\mathbf{B}$, where $F^{\mu \nu}$ and $F^{* \mu \nu}=\epsilon^{\mu\nu\alpha\beta}F_{\alpha\beta} / 2$  $(\epsilon^{0123} = 1)$ are the field strength tensor and its dual, respectively~\citep{1951PhRv...82..664S}. The aligned vacuum dipole  model for a neutron star with a magnetic field $B_{\rm p}$ at the pole and angular frequency $\Omega$ has a magnetic field and an induced electric field; the dipole field with magnetic axis along the z-direction of rotation axis and the induced electric field are
\begin{eqnarray}
\vec{B} &=& \frac{B_{\rm p}}{2} \Bigl(\frac{R}{r} \Bigr)^3 \Bigl[2\cos \theta \hat{r} + \sin \theta \hat{\theta} \Bigr], \\
\vec{E} &=& - \frac{B_{\rm p}}{2} \Bigl(\frac{R}{r} \Bigr)^4 \Bigl(\frac{R \Omega}{c} \Bigr) \Bigl[\frac{3\cos 2\theta + 1}{2} \hat{r} + \sin 2 \theta \hat{\theta} \Bigr],     
\end{eqnarray}
where $\hat{r}$ and $\hat{\theta}$ are unit vectors in the directions of $r$ and $\theta$ in spherical coordinates, and $R$ is the radius of the neutron star and $\theta$ is the angle from the north pole. The magnetic dipole moment is $B_{\rm p} R^3/2$. The electric field points toward the north and south poles and has a radial component inward when $\cos 2\theta > - 1/3$ and outward otherwise.
Then, the Maxwell scalar and pseudo-scalar in the laboratory frame are
\begin{eqnarray}
{\cal F} &=& \frac{B_{\rm p}^2}{8} \Bigl(\frac{R}{r} \Bigr)^6 \Biggl[1+ 3 \cos^2 \theta \nonumber\\
&&- \Bigl(\frac{R}{r} \Bigr)^2 \Bigl(\frac{R \Omega}{c} \Bigr)^2 \Bigl(1 - 2 \cos^2 \theta + 5 \cos^4 \theta \Bigr) \Biggr], \\
{\cal G} &=& B_{\rm p}^2 \Bigl(\frac{R}{r} \Bigr)^7 \Bigl(\frac{R \Omega}{c} \Bigr) \cos^3 \theta.
\end{eqnarray}
Millisecond neutron stars satisfy $R\Omega/c = 2\pi R/(cP) \simeq 0.252R_{12}/P_{-3}$, well below unity for the fiducial case, and are further constrained by the mass-shedding limit $R\Omega/c \lesssim \sqrt{GM/(c^2R)}$, where $R_{12}=R/(12\,{\rm km})$ and $P_{-3}=P/(10^{-3}\,{\rm s})$. 

\begin{figure}[!htbp]
\centering
\includegraphics[width=1\linewidth]{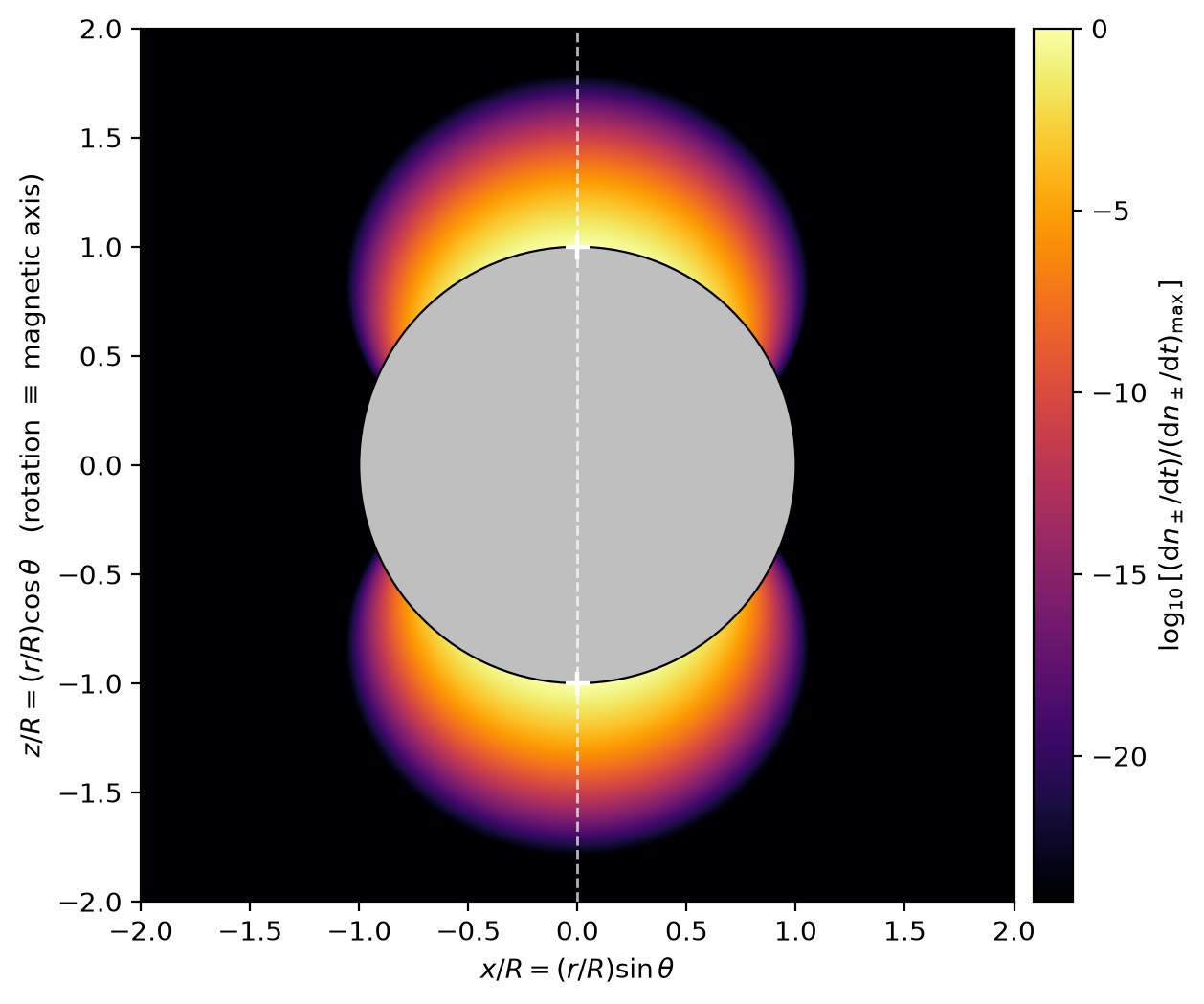} 
\caption{Density map of the pair creation rate via the Schwinger process outside the magnetar surface, evaluated as an unscreened aligned-vacuum estimate. The lower panel shows a zoom-in of the polar region from the upper panel. The density is normalized to its maximum value, $\mathrm{d}n_{\pm}/\mathrm{d}t$ at $(r = R, \theta = 0)$. In this figure, we adopt $R = 12\,\mathrm{km}$, $B_{\rm p} = 10^{14}\,\mathrm{G}$, and $P = 1\,\mathrm{ms}$. The corresponding maximum value is $\mathrm{d}n_{\pm}/\mathrm{d}t (r = R, \theta = 0) = 1.78 \times 10^{48}\,\mathrm{cm^{-3}\,s^{-1}}$. 
} 
\label{fig:pairmap}
\end{figure}

The Schwinger process is the pair creation of electrons and positrons by strong electric fields. (When the electric field is $E = (m/m_{\rm e})^2 E_{\rm Q}$, a charge with mass $m$ can also be produced by the Schwinger process. In standard particle physics, the electron and positron are the lightest charged particle and antiparticle.) In fact, one pair can be produced by an electric field comparable to or above the critical field per unit Compton volume and time. This can be understood by Sauter-type electric fields that are localized in space or time $E(z) = E_0 / \cosh^2 (z/\ell)$ or $E(t) = E_0 / \cosh^2 (t/\tau)$. When $eE_0 \ell \geq m c^2$ or $eE_0 c \tau \geq mc^2$, a pair is roughly produced in Compton volume and time~\citep{nikishov1970barrier,kim2002schwinger,kim2007improved}. 
The critical electric field has an energy in one Compton volume equal to about the rest mass energy of 2.7 electron-positron pairs.
This holds when a magnetic field has a component parallel to an electric field. Then the pair creation rate for electrons and positrons per unit volume is given by
\begin{align}
\frac{d n_{\pm}}{dt} = \frac{e^2 {\cal B} {\cal E}}{4 \pi^2 \hbar^2 c} \coth \Bigl(\pi \frac{{\cal B}}{{\cal E}} \Bigr) e^{- \pi m_{\rm e}^2 c^3/ \hbar e {\cal E}},
\end{align}
where $e$ is the value of the electron charge, and
\begin{eqnarray}
{\cal B} = \sqrt{\sqrt{{\cal F}^2 + {\cal G}^2} + {\cal F}}, \quad {\cal E} = \sqrt{\sqrt{{\cal F}^2 + {\cal G}^2} - {\cal F}}. 
\label{eq:field_invariants}
\end{eqnarray}
When the electric field is parallel or antiparallel to the magnetic field, ${\cal B} = |{\mathbf B}|$ and ${\cal E} = |{\mathbf E}|$.  When the two fields are perpendicular to each other (${\cal G}=0$),  ${\cal B}=0$ if $\mathbf{E}^2 \ge \mathbf{B}^2$, and ${\cal E}=0$ if $\mathbf{E}^2 \le \mathbf{B}^2$.
Electrons and positrons are produced by equal numbers, and the plasma of pairs is charge neutral. 
We may rewrite the pair creation rate in terms of quantities normalized to Compton scales as
\begin{align}
\frac{d n_{\pm}}{dt} &= \left[ \frac{\bar{\cal B} \bar{\cal E}}{4 \pi^2}
\coth \Bigl(\pi \frac{\bar{\cal B}}{\bar{\cal E}} \Bigr)
e^{- \pi / \bar{\cal E}} \right]
\Bigl( \frac{1}{\lambda_{\rm C}} \Bigr)^3
\Bigl(\frac{c}{\lambda_{\rm C}} \Bigr).
\end{align}
where $\bar{\cal B} = {\cal B}/B_{\rm Q}$ and $\bar{\cal E} = {\cal E}/E_{\rm Q}$ are dimensionless invariants and $\lambda_{\rm C} = \hbar/(m_{\rm e} c)$ is the Compton wavelength \cite{1951PhRv...82..664S,2010PhR...487....1R}; $\left[\cdots \right]$  is the number of pairs generated per unit Compton spacetime volume. For the reference condition of $R = 12 \,{\rm km}$ and $P = 10^{-3}\,{\rm s}$, $R \Omega /c = 0.252$, $|{\cal G}/{\cal F}|<0.54$, and ${\cal B}/{\cal E}\simeq 2 |{\cal F}/{\cal G}|+|{\cal G}/{\cal F}|/2-|{\cal G}/{\cal F}|^3 /8>3.95$; $\coth \bigl(\pi \bar{\cal B}/\bar{\cal E} \bigr) = 1 + 2 \sum_{k = 1}^{\infty} e^{- 2 \pi k \bar{\cal B}/\bar{\cal E} }$ can be safely replaced by 1. Then, the total number of electron-positron pairs generated per unit time is given by
\begin{align}
\dot N_{\pm}  = 2 \pi  \Bigl( \frac{1}{\lambda_{\rm C}} \Bigr)^3
\Bigl(\frac{c}{\lambda_{\rm C}} \Bigr)
\int_{0}^{\pi}  d \theta \sin \theta
\int_{R}^{\infty} dr \, r^2 \frac{\bar{\cal B} \bar{\cal E}}{4 \pi^2}
e^{- \pi / \bar{\cal E}} .
\label{eq:number_rate}
\end{align}
As shown in Figure \ref{fig:pairmap}, pairs are predominantly produced near the poles. 
Near the poles ${\cal G}>0$, so ${\bf E}$ is antiparallel to ${\bf B}$ and the radial electric field points inward: created electrons are accelerated outward along the open polar field, while positrons are driven back toward the surface. Since the Schwinger process creates electrons and positrons in equal numbers, the star plus magnetosphere remains globally charge neutral, although the two species are spatially separated.  
  
   When  $|{\cal G}/{\cal F}|<1$, we may simplify $\cal E$  and $\cal B$ by taking the leading orders of the expansions ${\cal B}=\sqrt{2{\cal F}}(1+ |{\cal G}/{\cal F}|^2/8-5|{\cal G}/{\cal F}|^4/128+\cdots)$ and ${\cal E}=\sqrt{2{\cal F}}(|{\cal G}/{\cal F}|/2-|{\cal G}/{\cal F}|^3/16+\cdots)$: ${\cal B}\simeq\sqrt{2{\cal F}}$ and  ${\cal E}\simeq|G|/\sqrt{2{\cal F}}$. The relative ~error of this approximation is smaller than 0.033 when $|{\cal G}/{\cal F}|<0.54$.  Upon changing the variable $y= \cos \theta$ and defining the dimensionless parameters,
\begin{equation}
\chi \equiv \beta \epsilon_{\rm \Omega}, \qquad
\beta \equiv \frac{B_{\rm p}}{B_{\rm Q}}, \qquad
\epsilon_{\rm \Omega} \equiv \frac{R\Omega}{c},
\end{equation}
the radial integration can be done under the assumption of $\epsilon_\Omega^2\ll1$ into the incomplete Gamma function~\citep{olver2010nist}, and Equation~(\ref{eq:number_rate}) becomes
\begin{equation}
\dot N_{\pm} =
\frac{cR^3}{8 \lambda_{\rm C}^4}\,\beta
\int_0^1 dy\, \sqrt{1+3y^2}\,
\Gamma\!\left[-1,\zeta(y)\right],
\label{eq:Ndot_exact}
\end{equation}
with
\begin{equation}
\zeta(y) =
\frac{\pi}{2\chi}\,\frac{\sqrt{1+3y^2}}{y^3}.
\end{equation} 

The angular integration may be found by a fitting formula, and the total emission per second may take the form of
\begin{align}
\dot N_{\pm} &= \frac{c R^3}{4 \pi^2 \lambda_{\rm C}^4} \beta \chi
e^{-\frac{\pi}{\chi} } f(\frac{\pi}{2\chi}).
\label{eq:sp-rate-approximation}
\end{align}
where
\begin{align}
f(\frac{\pi}{2\chi}) &= \frac{Q_4^2}{(\frac{3\pi}{2\chi} Q_4 + P_4)^2}.
\end{align}
Here, $P_4$ and $Q_4$ are Pad\'{e} approximants of the fourth order:
\begin{align}
P_4 &= 2 - 1.20920221 \left(\frac{\pi}{2\chi}\right)^{1/3} + 6.49881706 \left(\frac{\pi}{2\chi}\right)^{2/3} \nonumber\\
&\quad - 4.07067873 \frac{\pi}{2\chi} + 2.43121324 \left(\frac{\pi}{2\chi}\right)^{4/3}, \nonumber\\
Q_4 &= 1 - 0.56264565 \left(\frac{\pi}{2\chi}\right)^{1/3} + 1.66167994 \left(\frac{\pi}{2\chi}\right)^{2/3} \nonumber\\
&\quad - 0.95634061 \frac{\pi}{2\chi} + 0.56091913 \left(\frac{\pi}{2\chi}\right)^{4/3}.
\end{align}
This fitting formula is accurate to within 0.1\% in the wide range $10^{-5} \leq \frac{\pi}{2\chi} \leq 10^5$.
The most important factor in the Schwinger process by compact objects is $e^{-\pi/\chi}$, which regulates or smoothly turns on and off the pair creation: the Schwinger process is efficient when $\chi \geq 1$ but exponentially suppressed when $\chi \ll 1$. More detailed asymptotic expansions of the pair creation rate, along with the corresponding physical explanations, are provided in the \ref{sec:asymptotics}.

Electrons and positrons produced by the Schwinger process are accelerated in opposite directions by the induced electric field and produce a polarization current on a short time scale. In a self-consistent magnetosphere this current enters Maxwell's equation and reduces the parallel  electric field, so the vacuum solution applies only before the discharge loads the field with plasma. The rate derived above is therefore the initial unscreened rate, or an upper bound on the local pair-creation rate before screening and current closure. The energy-loss estimate below asks what would follow in the limiting case where the unscreened field is continuously replenished by the rotating magnetic field, with the energy ultimately drawn from the rotational reservoir. Compared with the total magnetic field energy inside and outside of the aligned vacuum dipole model
\begin{eqnarray}
E_{\rm em} = \frac{1}{12} B_{\rm p}^2 R^3 \Bigl( 1+ \frac{3}{10} \epsilon_{\rm \Omega}^2 \Bigr)
\end{eqnarray}
the vacuum-limit pair creation rate~(\ref{eq:sp-rate-approximation}) contains the large factor $R\Omega/\lambda_{\rm C}^4 B_{\rm Q}^2$. A sustained unscreened discharge would therefore be an efficient spin-powered channel, but only if the field were restored after pair backreaction. As the rotation of a millisecond magnetar slows, the unscreened pair creation rate is exponentially suppressed through $\exp[-\pi/(\beta\epsilon_{\Omega})]$. For example, the pair creation rate is exponentially suppressed by $e^{-\pi/\chi} = 2.67 \times 10^{-109}$ even for the magnetar with $B_{\rm p} = 50 B_{\rm Q}$ and a period of 1 second. Most magnetars in the McGill catalog have magnetic fields above $B_Q$ but their periods are seconds~\citep{2014ApJS..212....6O}, and even the very young radio-loud magnetar Swift~J1818.0-1607 already has $P=1.36\,\mathrm{s}$ at a characteristic age of only $240\,\mathrm{yr}$ \citep{2020ApJ...896L..30E}.

The total number of electron-positron pairs per unit time multiplied by their rest mass energy defines a vacuum upper-bound luminosity, $L_{\rm Sp}^{\rm vac} = 2 m_{\rm e} c^2 \dot N_{\pm}$. After screening is included, this term is multiplied by the duty factor defined in Section~\ref{sec:screening}; if the vacuum rate were maintained with unit duty cycle, it would give an upper-bound channel for the loss of rotational energy,
\begin{eqnarray}
-\dot{E}_{\rm rot} = L_{\rm Sp}^{\rm vac}(\epsilon_{\rm \Omega}, \beta).
\end{eqnarray}
The following formulae use this vacuum-limit convention. The Schwinger process depends only on two dimensionless quantities: the ratio of the surface velocity to the speed of light and another ratio of the pole magnetic field to the critical field. The rate of angular frequency change
\begin{eqnarray}
-\frac{\dot{\Omega}}{\Omega} &=& 2 \frac{m_{\rm e} c^2}{I \Omega^2} \dot N_{\pm} \nonumber\\
\end{eqnarray}
is approximately for $\chi \gg 1$ (extremely strong magnetic field and/or fast rotation)
\begin{eqnarray}
-\frac{\dot{\Omega}}{\Omega} \simeq \frac{1}{8 \pi^2} \frac{m_e c^2}{I \Omega} \Bigl(\frac{R}{\lambda_{\rm C}} \Bigr)^4
\Bigl(\frac{B_p}{B_Q} \Bigr)^{2} e^{- \pi \frac{B_{\rm Q}}{B_{\rm p}} \frac{c}{R\Omega}},
\label{big-x}
\end{eqnarray}
and for $\chi \ll 1$ (weak magnetic field and/or slow rotation)
\begin{eqnarray}
-\frac{\dot{\Omega}}{\Omega} \simeq \frac{2}{9 \pi^4} \frac{m_{\rm e} c^2}{I} \frac{R}{c} \Bigl(\frac{R}{\lambda_{\rm C}} \Bigr)^4
\Bigl(\frac{B_{\rm p}}{B_{\rm Q}} \Bigr)^{4} \Bigl(\frac{R\Omega}{c} \Bigr) e^{- \pi \frac{B_{\rm Q}}{B_{\rm p}} \frac{c}{R\Omega}}.
\end{eqnarray}
When the Schwinger process is a very efficient emission channel, that is, $\chi \gg 1$, Equation~(\ref{big-x}) can be solved in terms of the Lambert function~\citep{olver2010nist} as
\begin{eqnarray}
\Omega = - \frac{\pi (B_{\rm Q}/B_{\rm p})(c/R)}{2 W_{-1} \Bigl[ - \pi \sqrt{2 \pi \Bigl(\frac{I}{m_{\rm e} c^2} \Bigr) \Bigl( \frac{c}{R}\Bigr) 
\Bigl(\frac{\lambda_{\rm C}}{R} \Bigr)^4 \Bigl( \frac{B_{\rm Q}}{B_{\rm p}} \Bigr)^3 } \frac{1}{\sqrt{t}} \Bigr]}.
\end{eqnarray}

Magnetars also emit magnetic dipole radiation and gravitational waves~\citep{lu2018electromagnetic}. In Section~\ref{sec:spindown}, we combine these standard losses with the screened Schwinger term,
\begin{eqnarray}
-\dot{E}_{\rm rot} = L_{\rm dp} + L_{\rm gw} + L_{\rm Sp,eff}.   
\end{eqnarray}

\begin{figure}[!htbp]
\centering
\includegraphics[width=\columnwidth]{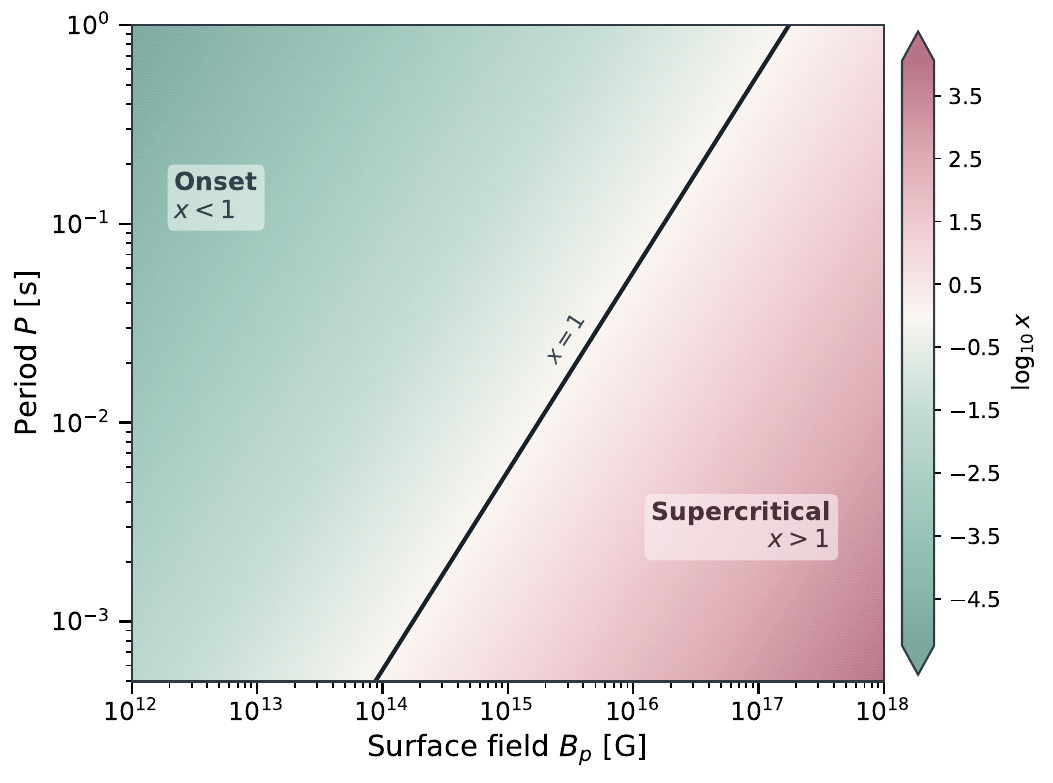}
\caption{Parameter-space map of the dimensionless combination $\chi=\beta\epsilon_{\rm \Omega}=(B_{\rm p}/B_{\rm Q})(2\pi R/cP)$ in the $(B_{\rm p},P)$ plane for $10^{12}\,\mathrm{G}\le B_{\rm p}\le10^{18}\,\mathrm{G}$ and $0.5\,\mathrm{ms}\le P\le1\,\mathrm{s}$. The solid line marks $\chi=1$, which separates the onset regime ($\chi<1$) from the supercritical regime ($\chi>1$) in the unscreened vacuum model. The color scale shows $\log_{10}x$, with deeper colors indicating larger departures from the transition.}
\label{fig:x_regime_map}
\end{figure}

\section{Screening estimate and range of validity of the vacuum rate}
\DIFaddbegin \label{sec:screening}
\DIFaddend 

The large vacuum rate in Equation~(\ref{eq:sp-rate-approximation}) requires at least a parametric treatment of pair backreaction. The starting point is the semiclassical Maxwell equation summarized in Appendix~\ref{sec:backreaction}: the created Dirac field contributes the expectation value of its current,
\begin{equation}
\partial_\alpha F^{\alpha\beta}=\frac{4\pi}{c}\langle J^\beta\rangle .
\label{eq:semiclassical_screening_current}
\end{equation}
Schwinger production creates electrons and positrons together, so the newly created plasma is initially charge-neutral, $\langle J^0\rangle\simeq0$. Screening is then supplied by charge separation and by the conduction current: the two species accelerate in opposite directions and contribute with the same sign to the current along the local electric field. Once this current develops, the exterior field no longer satisfies the vacuum Maxwell equations. In three-vector form,
\begin{equation}
\nabla\times{\bf B}-\frac{1}{c}\frac{\partial{\bf E}}{\partial t}
=\frac{4\pi}{c}{\bf J},
\label{eq:maxwell_current}
\end{equation}
and the current component along the magnetic field reduces $E_{\parallel}$. Locally, the parallel field evolves schematically as
\begin{equation}
\frac{\partial E_{\parallel}}{\partial t}
\simeq -4\pi\left(J_{\parallel}-J_{\rm ext}\right),
\label{eq:epar_screening}
\end{equation}
where $J_{\rm ext}$ denotes the current required by the global magnetosphere. To connect this equation with the Schwinger source term, consider a short flux-tube element near the pole, where $E_{\parallel}$ and $B$ may be treated as uniform during the first discharge. Let ${\cal W}(E_{\parallel},B)$ denote the local invariant Schwinger production rate per unit volume. The QED current can then be written, to the accuracy needed here, as
\begin{equation}
\langle J^0\rangle\simeq0,\qquad
\langle J_{\parallel}(t)\rangle
=J_{\rm pol}(t)+J_{\rm cond}(t),
\label{eq:qed_current_split}
\end{equation}
with
\begin{equation}
J_{\rm cond}(t)\simeq
2ec\int_0^t{\cal W}\!\left[E_{\parallel}(t'),B\right]dt',
\qquad
J_{\rm pol}E_{\parallel}\sim2m_{\rm e}c^2{\cal W}.
\label{eq:qed_current_terms}
\end{equation}
The first relation states that pair creation is charge symmetric, while the second separates the polarization current associated with pair creation from the conduction current of the accelerated pairs. The factor of two in $J_{\rm cond}$ counts the electron and positron streams, which move in opposite directions but carry the same sign of current. At the surface pole of the aligned solution,
$E_{\parallel,*}\simeq (R\Omega/c)B_{\rm p}=\chi B_{\rm Q}$. The acceleration time to relativistic motion is
\begin{equation}
t_{\rm acc}\sim\frac{m_{\rm e}c}{eE_{\parallel,*}}
=\frac{\lambda_{\rm C}}{c}\frac{B_{\rm Q}}{E_{\parallel,*}}
\simeq2.3\times10^{-21}
\left(\frac{E_{\parallel,*}}{0.57B_{\rm Q}}\right)^{-1}\,\mathrm{s}.
\label{eq:tacc_screening}
\end{equation}
Equivalently, $J_{\rm cond}/J_{\rm pol}\sim t/t_{\rm acc}$. For the fiducial polar field, $t_{\rm acc}$ is much shorter than the field-collapse and plasma times obtained below, so the screening estimate is controlled by the conduction current. During the first discharge we keep ${\cal W}$ fixed at its unscreened value ${\cal W}_*\equiv\dot n_{\pm,*}$, obtaining
\begin{equation}
\langle J_{\parallel}\rangle\simeq2ec\,\dot n_{\pm,*}t,\qquad
|\Delta E_{\parallel}|\simeq4\pi e c\,\dot n_{\pm,*}t^2.
\label{eq:early_current_growth}
\end{equation}
This is the local reduction of the semiclassical QED current to the newborn-magnetar parameters used below. The vacuum calculation therefore describes only the pre-screening state, before the created pairs supply the current needed to reduce $E_{\parallel}$.

The active height can be estimated from the same unscreened field rather than introduced as a free length. Along the magnetic axis, $B=B_{\rm p}(R/r)^3$ and $E_{\parallel}=E_{\parallel,*}(R/r)^4$. Keeping the leading Schwinger factor gives
\begin{equation}
\frac{{\cal W}(r)}{{\cal W}_*}
\simeq
\rho^{-7}\exp\left[-\frac{\pi}{\chi}\left(\rho^4-1\right)\right],
\qquad
\rho\equiv\frac{r}{R},
\label{eq:rate_height_profile}
\end{equation}
where ${\cal W}_*={\cal W}(R)$ and $\chi=E_{\parallel,*}/B_{\rm Q}$. We define the effective Schwinger height as the local logarithmic scale height of this rate,
\begin{equation}
\ell_{\rm Sch}\equiv
\left[-\frac{\partial \ln{\cal W}}{\partial r}\right]_{R,\,\theta=0}^{-1}
=\frac{R}{7+4\pi/\chi}.
\label{eq:ell_sch}
\end{equation}
For the fiducial model, $B_{\rm p}=10^{14}\,\mathrm{G}$, $R=12\,\mathrm{km}$, and $P=1\,\mathrm{ms}$ give $\chi\simeq0.57$ and $\ell_{\rm Sch}\simeq4.13\times10^4\,\mathrm{cm}$. This height is used below for the screening estimate.

For an active layer of length $\ell_{\rm Sch}$ along the field, the one-sign charge density needed to produce an electrostatic field comparable to the unscreened parallel field is
\begin{equation}
n_{\rm scr}\sim \frac{E_{\parallel}}{4\pi e\ell_{\rm Sch}}.
\label{eq:nscr}
\end{equation}
For the fiducial model this gives
\begin{equation}
n_{\rm scr}
\simeq 1.0\times10^{17}
\left(\frac{E_{\parallel,*}}{0.57B_{\rm Q}}\right)
\left(\frac{\ell_{\rm Sch}}{4.13\times10^4\,\mathrm{cm}}\right)^{-1}
\,\mathrm{cm^{-3}}.
\label{eq:nscr_num}
\end{equation}
Using the local vacuum rate quoted in Figure~\ref{fig:pairmap},
$\dot n_{\pm,*}\simeq1.78\times10^{48}\,\mathrm{cm^{-3}\,s^{-1}}$, the charge-filling time is
\begin{equation}
t_{\rm fill}\sim \frac{n_{\rm scr}}{\dot n_{\pm,*}}
\simeq 5.6\times10^{-32}
\left(\frac{\ell_{\rm Sch}}{4.13\times10^4\,\mathrm{cm}}\right)^{-1}
\,\mathrm{s}.
\label{eq:tfill}
\end{equation}
This is not the discharge duration; it only shows that pair production supplies the required charge density on a time far below any macroscopic scale once the local field reaches the Schwinger regime.

Two other times enter the field oscillation estimate. Setting $|\Delta E_{\parallel}|\sim E_{\parallel,*}$ in Equation~(\ref{eq:early_current_growth}) gives the Ampere-law time
\begin{equation}
t_A\sim
\left(\frac{E_{\parallel,*}}{4\pi e c\dot n_{\pm,*}}\right)^{1/2}
\simeq 2.8\times10^{-19}\,\mathrm{s}.
\label{eq:tampere}
\end{equation}
The collective plasma response time for a pair plasma with density of order $n_{\rm scr}$ is
\begin{equation}
\omega_{\rm p}^{-1}
\sim
\left(\frac{m_{\rm e}}{8\pi e^2 n_{\rm scr}}\right)^{1/2}
\simeq
3.9\times10^{-14}
\left(\frac{n_{\rm scr}}{1.0\times10^{17}\,\mathrm{cm^{-3}}}\right)^{-1/2}
\mathrm{s}.
\label{eq:wp_screening}
\end{equation}
We therefore define the local screening time as
\begin{equation}
t_{\rm scr}\equiv
\max\left(t_{\rm fill},\,t_A,\,\omega_{\rm p}^{-1}\right).
\label{eq:tscr_definition}
\end{equation}
For the fiducial parameters, $t_{\rm scr}\simeq\omega_{\rm p}^{-1}\simeq3.9\times10^{-14}\,$s, and the time ordering is
\begin{equation}
\begin{aligned}
t_{\rm fill}&\ll t_A\ll t_{\rm scr}\ll
\frac{\ell_{\rm Sch}}{c}\simeq1.38\times10^{-6}\,\mathrm{s},\\
\frac{\ell_{\rm Sch}}{c}&\ll \frac{R_{\rm LC}}{c}=\Omega^{-1}
\simeq1.59\times10^{-4}\,\mathrm{s}
\ll P=10^{-3}\,\mathrm{s}.
\end{aligned}
\label{eq:screening_hierarchy}
\end{equation}

A sustained Schwinger luminosity requires the unscreened field to be restored after this local collapse. We parameterize that global reopening by a replenishment time $t_{\rm rep}$ and a duty factor
\begin{equation}
f_{\rm scr}=\min\left(1,\frac{t_{\rm scr}}{t_{\rm rep}}\right),\qquad
\dot N_{\pm,\rm eff}=f_{\rm scr}\dot N_{\pm}^{\rm vac},
\qquad
L_{\rm Sp,eff}=f_{\rm scr}L_{\rm Sp}^{\rm vac}.
\label{eq:duty_screen}
\end{equation}
We use two reference choices below. A local light-crossing replenishment time, $t_{\rm rep}=\ell_{\rm Sch}/c$, gives
$f_{\ell/c}\simeq2.9\times10^{-8}$ and represents the fastest plausible reopening of a local Schwinger layer. For an open-field, spin-powered discharge, however, the voltage is maintained by the global unipolar-induction circuit. The causal time is then the Alfv\'en/light-cylinder crossing time
\begin{equation}
t_{\rm rep}=t_{\rm LC}\equiv\frac{R_{\rm LC}}{c}=\frac{1}{\Omega}=\frac{P}{2\pi},
\qquad
f_{\rm LC}\simeq2.5\times10^{-10}P_{-3}^{-1}.
\label{eq:tlc_replenish}
\end{equation}

This prescription does not replace a time-dependent kinetic or force-free calculation. It is sufficient, however, to keep the spin evolution from using a unit-duty-cycle vacuum rate. The quantities $\dot N_{\pm}^{\rm vac}$ and $L_{\rm Sp}^{\rm vac}$ remain useful as initial upper bounds, while $\dot N_{\pm,\rm eff}$ and $L_{\rm Sp,eff}$ provide the screened estimates used in Section~\ref{sec:spindown}.

\DIFaddend \section{Comparison with the Goldreich--Julian particle supply}
\label{sec:gj}

\begin{figure}[t]
\centering
\includegraphics[width=0.5\textwidth]{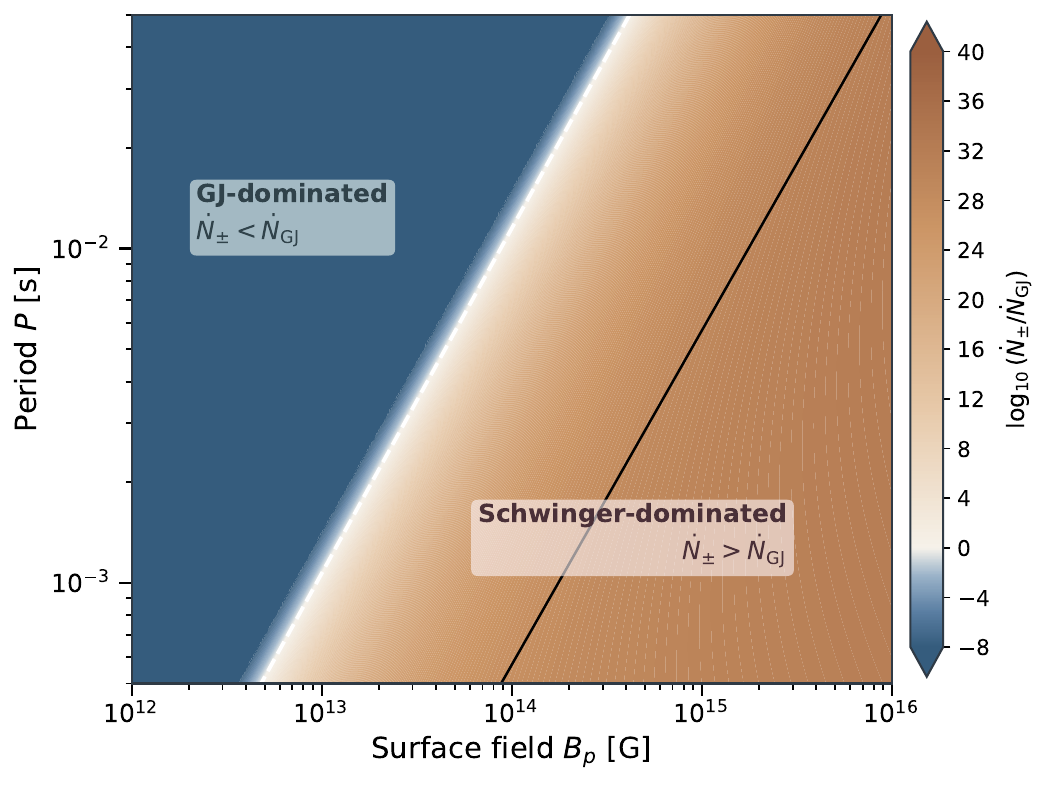}
\caption{Birth map in the $(B_{\rm p},P)$ plane, evaluated with the Pad\'{e} approximation to the unscreened vacuum Schwinger rate. The color scale shows $\log_{10}(\dot N_{\pm}/\dot N_{\rm GJ})$, where $\dot N_{\rm GJ}\approx B_{\rm p}R_{\rm NS}^3\Omega^2/(2ec)$. The comparison is a diagnostic of pair-creation capacity, not a direct estimate of the escaping open-field-line flux after screening. }
\label{fig:birth_maps}
\end{figure}

The usual Goldreich--Julian charge density is $\rho_{\rm GJ}\sim -\Omega B/(2\pi c)$ \citep{1969ApJ...157..869G}. On open field lines, the polar-cap area scales as
\begin{equation}
A_{\rm pc} \sim \pi R^2 \frac{R}{R_{\rm lc}}
\sim \pi \frac{R^3 \Omega}{c},
\end{equation}
so the classical open-field particle supply is approximately
\begin{equation}
\begin{aligned}
\dot N_{\rm GJ}
&\sim \frac{|\rho_{\rm GJ}|}{e} c A_{\rm pc}
\sim \frac{B_{\rm p} R^3 \Omega^2}{2ec} \\
&\simeq 2.37\times10^{38}\, B_{14} \, P_{\rm ms}^{-2}
\Bigl(\frac{R}{R_{12}}\Bigr)^3 \, \mathrm{s^{-1}}.
\end{aligned}
\label{eq:Ngj}
\end{equation}
Equation~(\ref{eq:Ngj}) is accurate up to factors of order unity that depend on the geometry of the open field region. The comparison below keeps this distinction explicit: $\dot N_{\rm GJ}$ is an open-field-line supply rate, whereas $\dot N_{\pm}$ is a global near-surface production rate in an unscreened vacuum field. Pairs created in this region may screen the field locally, remain on closed field lines, or carry magnetospheric currents rather than escape to infinity. For the present purpose, the ratio measures charge-production capacity before screening rather than an escaping flux.

For our fiducial newborn magnetar with $B_{\rm p}=10^{14}\,$G, $R=12\,$km, and $P_0=1\,$ms, the Pad\'{e} approximation gives
\begin{equation}
\dot N_{\pm} \simeq 2.13\times10^{64}\,\mathrm{s^{-1}}, \qquad
\dot N_{\rm GJ} \simeq 2.37\times10^{38}\,\mathrm{s^{-1}},
\end{equation}
so that
\begin{equation}
\frac{\dot N_{\pm}}{\dot N_{\rm GJ}} \simeq 8.99\times10^{25}.
\end{equation}

This large ratio should not be interpreted as a statement that Schwinger creation dominates the particle supply at all magnetic colatitudes. In the same aligned vacuum field, the surface effective Schwinger parameter has the leading angular dependence
\begin{equation}
\chi_\theta\equiv\frac{{\cal E}(R,\theta)}{B_{\rm Q}}
\simeq
\chi\,\frac{2\cos^3\theta}{\sqrt{1+3\cos^2\theta}},
\qquad
\chi=\frac{B_{\rm p}}{B_{\rm Q}}\frac{R\Omega}{c}.
\label{eq:chi_theta}
\end{equation}
Up to algebraic prefactors, the local rate changes as
\begin{equation}
\frac{{\cal W}(\theta)}{{\cal W}(0)}
\sim
\exp\left[-\pi\left(\frac{1}{\chi_\theta}-\frac{1}{\chi}\right)\right].
\label{eq:angular_suppression}
\end{equation}
The exponential sensitivity confines efficient Schwinger production to the polar region where $\chi_\theta$ is largest. Away from this region, or after spin-down lowers $\chi$, ordinary Goldreich--Julian extraction and gap-driven cascades may again provide the dominant local particle supply. The comparison with $\dot N_{\rm GJ}$ is therefore a statement about the initial near-polar charge-production capacity of the unscreened benchmark, not about the angularly averaged escaping flux after screening.
Within the present aligned vacuum dipole model, the unscreened Schwinger production rate exceeds the classical Goldreich--Julian supply by many orders of magnitude. Once triggered, vacuum pair creation is therefore not a small correction to the local charge budget; determining how the pairs divide between screening charges, closed-field currents, and escaping particles requires a magnetospheric transport calculation.

A scan of the full $(B_{\rm p},P)$ plane shows that the contour $\dot N_{\pm}=\dot N_{\rm GJ}$ (Figure \ref{fig:birth_maps}) already appears at $x\sim0.05$ in the Pad\'{e} model, well before the formal onset boundary at $\chi=1$ (Figure \ref{fig:x_regime_map}). The Schwinger channel can therefore dominate the initial local pair supply  even while the system remains in the near threshold tunneling regime. A newborn magnetar does not need to be deeply supercritical for the unscreened vacuum rate to outpace the classical Goldreich--Julian current scale.

\section{Misaligned Magnetars}
\label{sec:misaligned}
We now consider magnetars and highly magnetized neutron stars of which magnetics axes are not aligned with the rotation axes by an angle of $\alpha$.  The corresponding unscreened field is called the Deutsch field~\citep{deutsch1955electromagnetic,Michel_Li_1999,Satherley_Gordon_2022}, which reduces to the following form near the star surface, where $r\Omega/c \ll1$:  
\begin{eqnarray}
 \vec{B} &=& \frac{B_p}{2} \Bigl(
 \frac{R}{r}\Bigr)^3 \Bigl[2 \bigl( \cos \alpha \cos \theta + \sin \alpha \sin \theta \cos \varphi_s \bigr) \hat{r} \nonumber\\
 &&+ \bigl(\cos \alpha \sin \theta - \sin \alpha \cos \theta \cos \varphi_s \bigr) \hat{\theta} \nonumber\\
 &&+  \sin \alpha \sin \varphi_s \hat{\varphi}\Bigr],   
\end{eqnarray}
\begin{eqnarray}
 \vec{E} &=& - \frac{B_p}{2} \Bigl(
 \frac{R}{r}\Bigr)^4 \Bigl(\frac{R\Omega}{c} \Bigr) \Bigl[ \frac{1}{2} \Bigl( \cos \alpha \bigl( 3 \cos 2 \theta + 1 \bigr) \nonumber\\ && + 3 \sin \alpha \sin 2 \theta \cos \varphi_s \Bigr) \hat{r} \nonumber\\
 && + \Bigl( \cos \alpha \sin 2 \theta + \sin \alpha  \bigl( \frac{r^2}{R^2}-  \cos 2 \theta \bigr) \cos \varphi_s \Bigr) \hat{\theta} \nonumber\\
 && -  \bigl( \frac{r^2}{R^2}- 1 \bigr) \sin \alpha \cos \theta \sin \varphi_s \hat{\varphi}\Bigr]. 
\end{eqnarray}

\begin{figure}[t]
\centering
\includegraphics[width=0.5\textwidth]{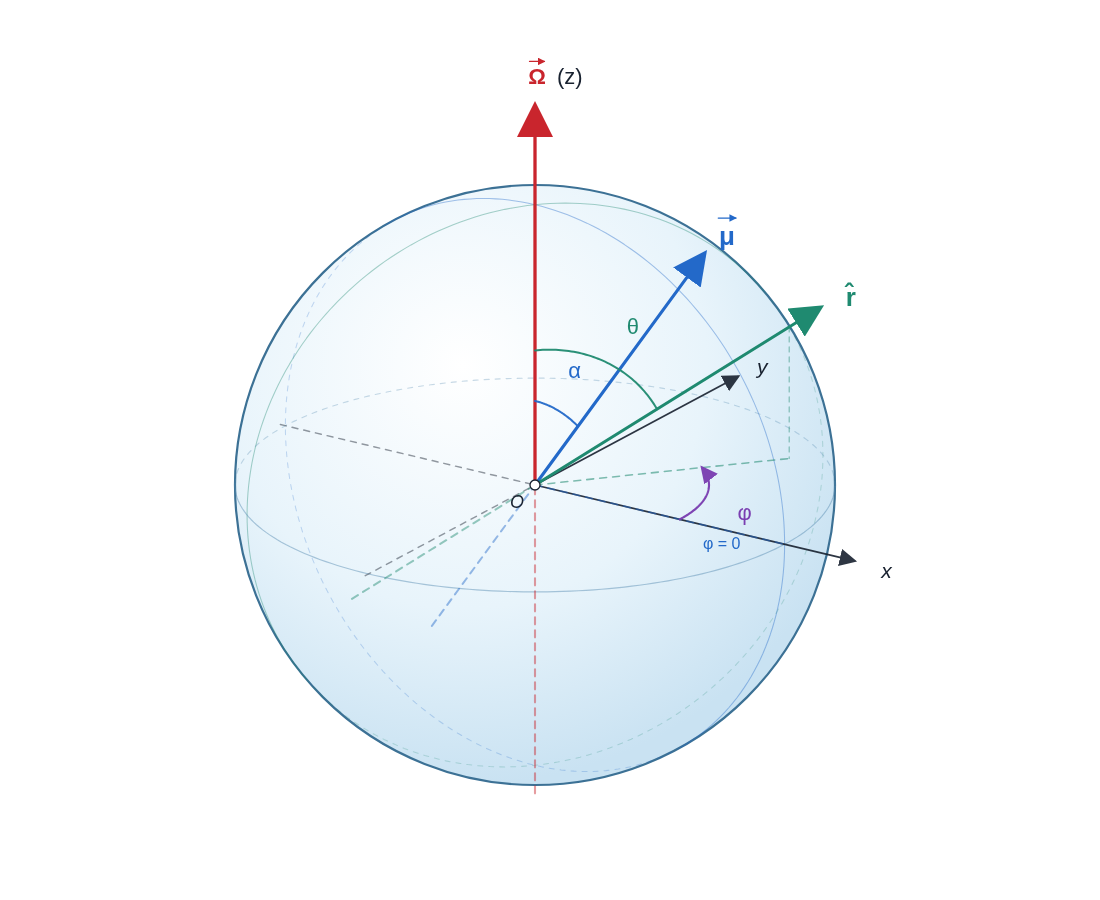}
\caption{Spherical geometry and angular definitions of the misaligned magnetar: $\vec{\mu}$ denotes the magnetic dipole and $\hat{\mu}$ is the unit vector along the magnetic pole.}
\label{fig:spherical_angles_definition}
\end{figure}

\begin{figure*}[t]
\centering
\includegraphics[width=\textwidth]{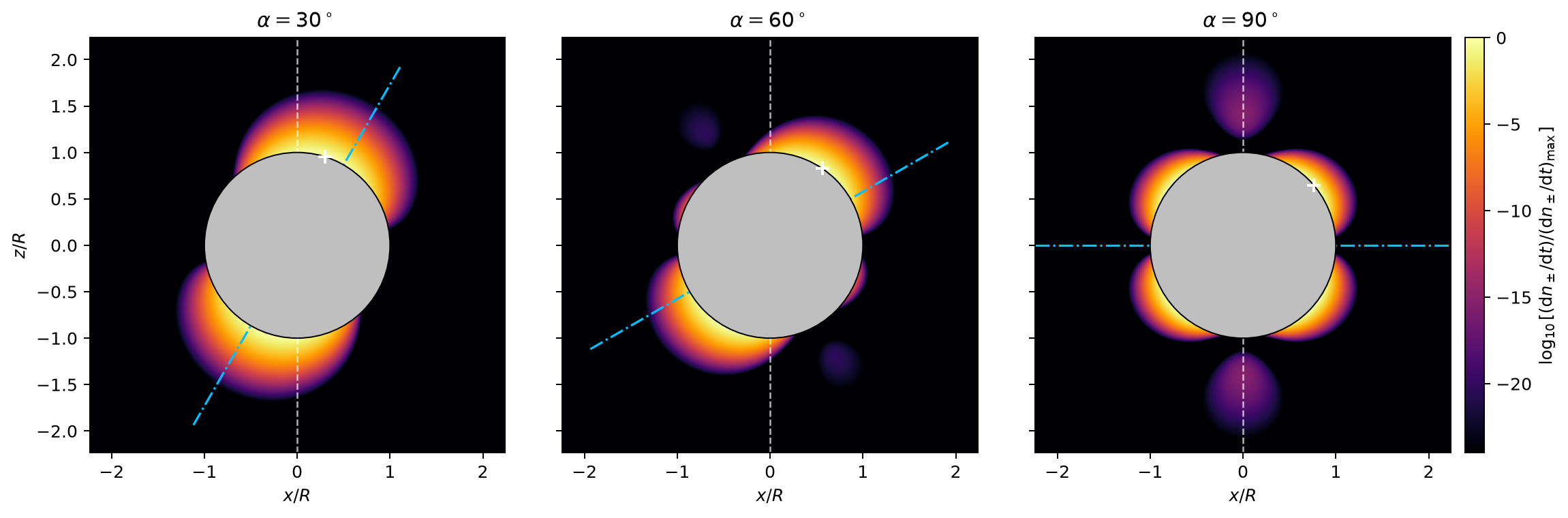}
\caption{Schwinger pair-creation rate for misaligned rotators with inclination $\alpha=30^\circ$, $60^\circ$, and $90^\circ$ (left to right), computed from the Deutsch near-zone field of Section~\ref{sec:misaligned} through the invariants ${\cal F}$, ${\cal G}$ and Equation (\ref{eq:field_invariants}), evaluated exactly, i.e. without expanding in $|{\cal G}/{\cal F}|$. Because an oblique rotator is neither axisymmetric nor equatorially symmetric, each panel shows the meridional plane that contains both the rotation axis and the magnetic axis: $x>0$ corresponds to $\varphi_s=0$ and $x<0$ to $\varphi_s=\pi$, so that the full angular range $0\le\theta\le2\pi$ in that plane is displayed. In this plane ${\cal T}={\cal A}_{\varphi}=0$. The white dashed line marks the rotation axis, the blue dot-dashed line the magnetic axis, and the white cross one of the symmetry-equivalent surface maxima. As in Figure~\ref{fig:pairmap}, we adopt $R=12\,\mathrm{km}$, $B_{\rm p}=10^{14}\,\mathrm{G}$, and $P=1\,\mathrm{ms}$
($\chi=0.570$), and each panel is normalized to its own surface maximum, $\mathrm{d}n_{\pm}/\mathrm{d}t|_{\max}=1.07\times10^{48}$, $1.66\times10^{47}$, and $1.29\times10^{45}\,\mathrm{cm^{-3}\,s^{-1}}$, i.e. $0.60$, $9.3\times10^{-2}$, and $7.3\times10^{-4}$ of the aligned value. Two features distinguish the oblique case from the aligned one. First, the
active region is not centered on the magnetic pole but is displaced toward the rotation axis, to $\theta=17.4^\circ$, $34.1^\circ$, and $50.0^\circ$
respectively, i.e. to magnetic colatitudes of $12.6^\circ$, $25.9^\circ$, and $40.0^\circ$. Second, at the magnetic pole the pseudo-scalar is exactly ${\cal G}=B_{\rm p}^{2}\epsilon_{\Omega}\cos\alpha$, so an orthogonal rotator has ${\cal G}=0$ along its magnetic axis and the polar cap switches off entirely; to leading order in $|{\cal G}/{\cal F}|$ this corresponds to $\bar{\cal E}\simeq\chi\cos\alpha$. The $\alpha=60^\circ$ panel is intermediate: the emission is still concentrated on one side of the magnetic axis, but the secondary lobes that develop into the four-lobe pattern of the orthogonal rotator are already visible. For $\alpha=90^\circ$ the emission indeed breaks into four lobes flanking the magnetic axis at a magnetic colatitude of $\simeq40^\circ$. The faint detached lobes near the rotation axis in the
$\alpha=60^\circ$ and $90^\circ$ panels originate from the inductive $r^{-2}$ part of the Deutsch field, which vanishes at the stellar surface; they lie some twenty orders of magnitude below the peak. The exponential factor $e^{-\pi/\bar{\cal E}}$ converts the modest geometric reduction of $\bar{\cal E}$ into a decrease of the peak rate by more than three orders of magnitude between $\alpha=0$ and $\alpha=90^\circ$. As in Figure~\ref{fig:pairmap}, these are unscreened vacuum estimates; the maps span $|x|,|z|\le\sqrt{5}\,R$, reaching $r\simeq3.2R$ at the corners, still inside the light cylinder at $R_{\rm LC}=3.98R$ where the near-zone reduction of the Deutsch field applies.}
\label{fig:pairmap_misaligned}
\end{figure*}

Here, as shown in Figure \ref{fig:spherical_angles_definition}, the rotation axis $\vec{\Omega}$ is along the z-axis, $\alpha$ is the inclination angle of the magnetic axis with respect to the rotation axis, $\theta$ the polar angle from the rotation axis, and $\varphi_s\equiv\varphi-\Omega t$ the longitude on the star; $\varphi_s=0$ refers to the longitude of the magnetic axis.

Introducing the following angular factors with $\eta\equiv r/R$, 
\begin{align}
{\cal C}&\equiv \cos\alpha\cos\theta+\sin\alpha\sin\theta\cos\varphi_s ,\\
{\cal S}&\equiv \cos\alpha\sin\theta-\sin\alpha\cos\theta\cos\varphi_s ,\\
{\cal T}&\equiv \sin\alpha\sin\varphi_s ,\\
{\cal A}_r&\equiv
\frac{1}{2}\left[\cos\alpha(3\cos2\theta+1)
 +3\sin\alpha\sin2\theta\cos\varphi_s\right],\\
{\cal A}_{\theta}&\equiv
\cos\alpha\sin2\theta
+\sin\alpha\left( \eta^2-\cos2\theta \right)\cos\varphi_s,\\
{\cal A}_{\varphi}&\equiv
-(\eta^2-1)\sin\alpha\cos\theta\sin\varphi_s,
\end{align}
one can express the Maxwell scalar and pseudo-scalar in a compact form: 
\begin{align}
{\cal F}&=\frac{B_{\rm p}^2}{8}\eta^{-6}
\left[ 4{\cal C}^2+{\cal S}^2+{\cal T}^2 -\eta^{-2}\epsilon_{\Omega}^2
\left({\cal A}_r^2+{\cal A}_{\theta}^2+{\cal A}_{\varphi}^2\right)
\right], \\
{\cal G}&=\frac{B_{\rm p}^2}{4}\eta^{-7}\epsilon_{\Omega}
\left(2{\cal C}{\cal A}_r+{\cal S}{\cal A}_{\theta}
+{\cal T}{\cal A}_{\varphi}\right).
\end{align}
The angular factors can be expressed as ${\cal C} = \cos \varpsi$ of the arc-length $\varpsi$ on a unit sphere between $\hat{r}$ and $\hat{\mu}$, and ${\cal S} = \sin \varpsi (\partial \varpsi/\partial \theta)$, 
${\cal T} = (\sin \varpsi/ \sin \theta) (\partial \varpsi/\partial \varphi_s)$.

When $\alpha = 0$, these expressions reduce to those of the aligned case in Section \ref{sec:model}. With these $\cal{F}$ and $\cal{G}$, the Schwinger pair production rate per unit volume $d n_\pm /dt$ can be calculated as in the aligned case, albeit a compact formula of the space-integrated production rate is unlikely to be obtained. In the co-rotating plane with $\varphi_s = 0$, the angular factors are simplified as
\begin{eqnarray}
{\cal C} &=& \cos (\theta - \alpha),\\
{\cal S} &=& \sin (\theta - \alpha),\\
{{\cal A}_r} &=& \frac{1}{2} \Bigl(3 \cos(2\theta - \alpha) + \cos \alpha \Bigr),\\
{{\cal A}_{\theta}} &=& \sin(2\theta - \alpha) + \eta^2 \sin \alpha,
\end{eqnarray}
with  ${\cal T}={\cal A_\varphi}=0$.
Numerical results for the Schwinger process are shown in Figure~\ref{fig:pairmap_misaligned} for $\alpha=30^\circ, 60^\circ, 90^\circ$. The characteristic features of the Schwinger process follow from the dominant factor $e^{- \pi \bar{\cal B}/|\bar{\cal G}|}$. The curves for zero pair production follow from $\bar{\cal G} = 0$,
\begin{eqnarray}
2{\cal C}{\cal A}_r+{\cal S}{\cal A}_{\theta}  = 0.   \end{eqnarray}
In the case of $\alpha = 90^\circ$, the condition for no pair production (zero-curve) is explicitly given by
\begin{eqnarray}
\cos \theta = 0, \quad \cos \theta = \pm \frac{\sqrt{5 - \eta^2}}{2}.    
\end{eqnarray}
One zero-curve is always $\theta = 90^\circ$ exterior magnetar, and the other curve is $\theta = 0^\circ, 180^\circ$  on the surface with $\eta =1$ but follows $\pm \arccos(\sqrt{5 - \eta^2}/2)$ in the region $\eta \leq \sqrt{5}$. In any region with $\eta > \sqrt{5}$, however, the Schwinger process turns on, which explains the two faint regions above (below) $\theta = 0^\circ \, (180^\circ)$ in Figure~\ref{fig:pairmap_misaligned}. The four dividing zero-curves result in the quadrupole figure of the Schwinger process in the domain of numerical results. 
Local maxima for pair production are approximately located at maxima of $|\bar{\cal G}|$, 
\begin{eqnarray}
\cos \theta \approx \pm \sqrt{\frac{5 - \eta^2}{12}},
\end{eqnarray}
which are $54.7^\circ, 125.3^\circ$, close to the exact numerical values $50.0^\circ, 130^\circ$. 
The zero curves and maxima explain the quadrupole figure of the Schwinger process for $\alpha = 90^\circ$. The maximum value of $d n_{\pm}/dt$ depends on the inclination angle $\alpha$, crucially through $|\bar{\cal G}|$, which decreases as $\alpha$ increases, and is numerically confirmed in Figure~\ref{fig:pairmap_misaligned}.

In addition to the Schwinger process that always produces pairs, regardless the inclination angle $\alpha$, the magnetic dipole radiates only for misaligned configurations with the luminosity~\citep{deutsch1955electromagnetic,Satherley_Gordon_2022}
\begin{align}
L_{\rm dp} = \frac{B_{\rm p}^2R^6\Omega^4\sin^2\alpha}{6c^3}.
\label{eq:dipole}
\end{align}
Thus misaligned magnetars lose rotational energy through dipole radiation as well as Schwinger emission, in strong contrast to Schwinger emission for aligned magnetars. Furthermore, misaligned magnetars have quadrupole moment from triaxial ellipsoidal figure due to both rotation and magnetic pressure, which is essential for gravitational waves in the next section.  

\section{Screened spin down with Schwinger, magnetic dipole, and gravitational wave losses}\label{sec:spindown}

\begin{figure*}[t]
\centering
\includegraphics[width=\textwidth]{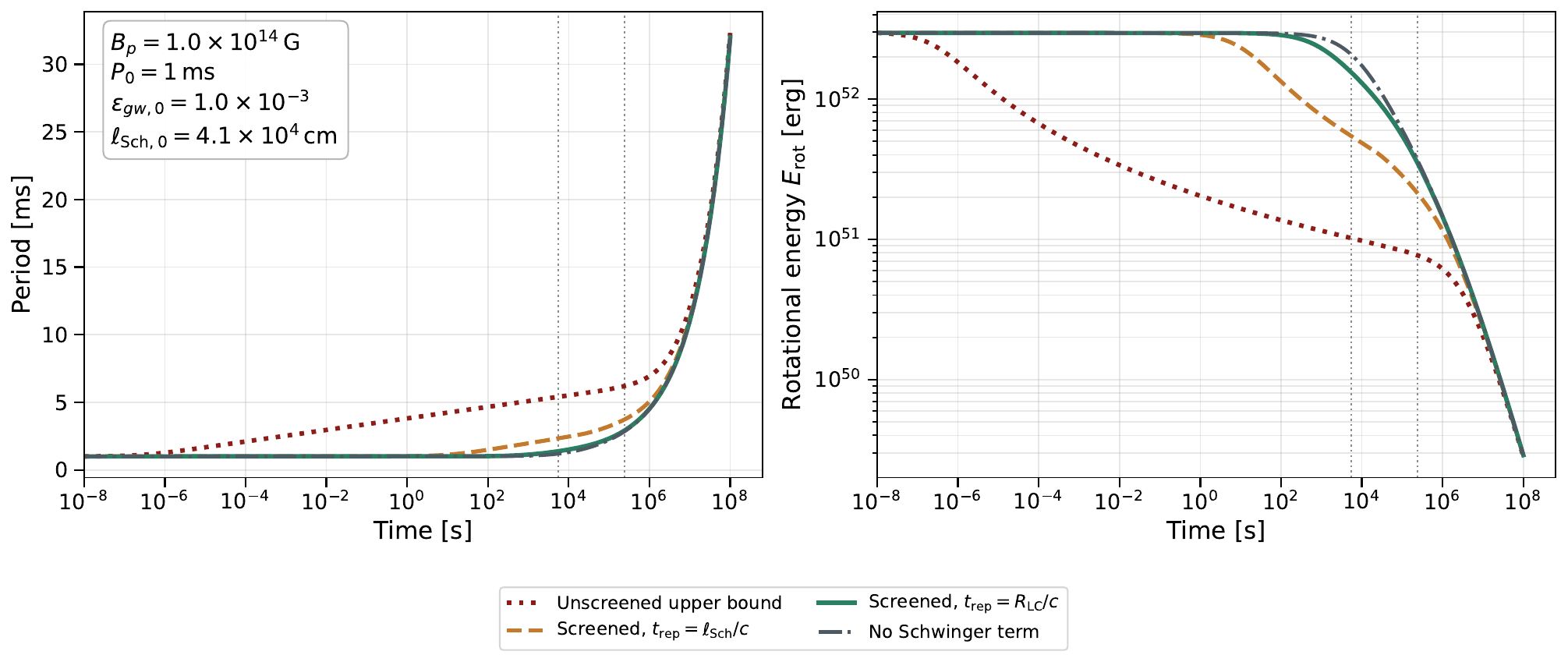}
\caption{Spin evolution of the fiducial magnetar model under four Schwinger prescriptions: the unit-duty-cycle unscreened upper bound, a screened local upper-limit model with $t_{\rm rep}=\ell_{\rm Sch}/c$, the fiducial global-circuit model with $t_{\rm rep}=R_{\rm LC}/c$, and a no-Schwinger reference. \emph{Left:} period as a function of time. \emph{Right:} rotational energy as a function of time. The vertical dotted lines mark $L_{\rm Sp,eff}=L_{\rm gw}$ and $L_{\rm gw}=L_{\rm dp}$ in the fiducial screened model. Screening moves the Schwinger-assisted spin drain from the microphysical discharge time to the replenishment time scale.}
\label{fig:period_energy}
\end{figure*}

\begin{figure*}[t]
\centering
\includegraphics[width=\textwidth]{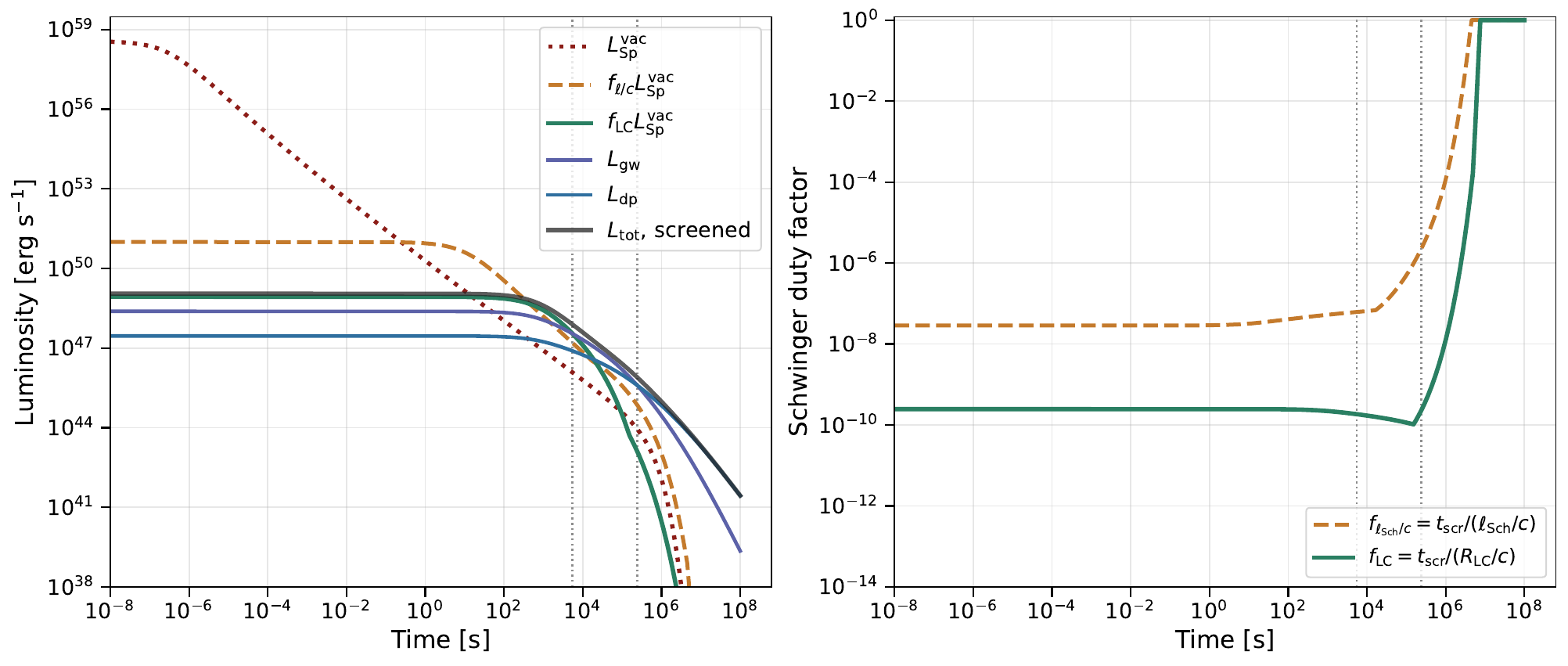}
\caption{Luminosity and duty-cycle evolution for the fiducial model. \emph{Left:} the unscreened Schwinger upper bound, the two screened Schwinger luminosities $f_{\ell_{\rm Sch}/c}L_{\rm Sp}^{\rm vac}$ and $f_{\rm LC} L_{\rm Sp}^{\rm vac}$, and the gravitational-wave and dipole luminosities. \emph{Right:} the corresponding duty factors $f_{\ell_{\rm Sch}/c}$ and $f_{\rm LC}$. In the fiducial global-circuit model, $f_{\rm LC}\simeq2.5\times10^{-10}$ at birth, so the sustained Schwinger luminosity is reduced from $L_{\rm Sp}^{\rm vac}\simeq3.5\times10^{58}\,\mathrm{erg\,s^{-1}}$ to $L_{\rm Sp,eff}\simeq8.6\times10^{48}\,\mathrm{erg\,s^{-1}}$.}
\label{fig:luminosities}
\end{figure*}

The rotational energy of the star is
\begin{align}
E_{\rm rot} = \frac{1}{2}I\Omega^2 = \frac{2\pi^2 I}{P^2},
\label{eq:erot}
\end{align}
where $P=2\pi/\Omega$ is the spin period. We adopt a screened Schwinger luminosity defined by the pair rest-mass flux,
\begin{align}
L_{\rm Sp}^{\rm vac} = 2m_{\rm e}c^2\dot N_{\pm}^{\rm vac},\qquad
L_{\rm Sp,eff}=f_{\rm scr}L_{\rm Sp}^{\rm vac}.
\label{eq:lsp}
\end{align}
This definition neglects any additional kinetic energy acquired by the pairs after creation and therefore gives a lower bound on the rest-mass energy associated with a specified pair rate. The factor $f_{\rm scr}$ applies the screening prescription of Equation~(\ref{eq:duty_screen}) and separates the sustained luminosity from the instantaneous vacuum upper bound.

The comparison in this section still combines two idealized geometries and should be used as a diagnostic. The Schwinger rate is computed for the aligned vacuum field, which provides a tractable reference for the initial unscreened $E_{\parallel}$. The dipole luminosity is the standard oblique-vacuum expression; in the fiducial example we set $\sin\alpha=1$ to show the largest conventional electromagnetic torque. A consistent oblique treatment would require the Deutsch field together with plasma screening and current closure. The geometrical change in the dipole term is expected to be of order unity, whereas the larger uncertainty in the Schwinger term is the replenishment duty cycle after local screening.

The screened Schwinger term and the dipole term should also not be read as two independent ways of radiating the same local electric field. Schwinger discharge acts first on the non-ideal component $E_\parallel$ and supplies charges that tend to push the near-surface magnetosphere toward a pair-loaded state. The dipole luminosity, by contrast, represents the global electromagnetic torque and Poynting flux of the rotating magnetic field. Pair loading can change the coefficient of this torque, for example by moving the magnetosphere from a vacuum-like state toward a force-free state, but it does not by itself eliminate magnetic-dipole spin-down. In this sense, a faster local pair-creation time does not imply that Schwinger energy loss must dominate the global dipole torque; dominance depends on the screened duty factor, the replenishment time, and the angular region in which the Schwinger threshold is reached.

The other two standard loss channels are the magnetic-dipole luminosity~(\ref{eq:dipole}) and the quadrupole gravitational-wave luminosity,
\begin{align}
L_{\rm gw} = \frac{32GI^2\epsilon_{\rm gw}^2\Omega^6}{5c^5},
\label{eq:lgw}
\end{align}
where $\alpha$ is the inclination angle and $\epsilon_{\rm gw}$ is the stellar ellipticity \citep{2018MNRAS.480.4402L}.
We treat $\epsilon_{\rm gw}$ as a phenomenological parameter. The fiducial value adopted below is within the range discussed for strongly magnetized neutron stars, although the true ellipticity depends on the internal magnetic geometry and on the field evolution.
The spin evolution is then governed by
\begin{align}
-I\Omega\dot\Omega = L_{\rm Sp,eff} + L_{\rm dp} + L_{\rm gw},
\label{eq:omegadot}
\end{align}
or, equivalently, by the period equation
\begin{align}
\dot P = \frac{P^3}{4\pi^2 I}\left(L_{\rm Sp,eff} + L_{\rm dp} + L_{\rm gw}\right).
\label{eq:pdot}
\end{align}
In the present calculation, we keep $B_p$, $\alpha$, and the stellar ellipticity constant during the evolution, adopting the fiducial choice
\begin{align}
\epsilon_{\rm gw}(P) = \epsilon_{{\rm gw},0} = 10^{-3},
\label{eq:eps_evolution}
\end{align}
which is consistent with estimates of magnetically induced ellipticities of order $10^{-3}$ in strongly magnetized neutron stars \citep{2022A&A...666A.138L}. We then integrate Equation~(\ref{eq:pdot}) numerically using Equation~(\ref{eq:sp-rate-approximation}) for $\dot N_{\pm}^{\rm vac}$ and Equation~(\ref{eq:duty_screen}) for $f_{\rm scr}$.

For the fiducial model, we adopt $R=12\,\mathrm{km}$, $I=1.5\times10^{45}\,\mathrm{g\,cm^2}$, $B_p=10^{14}\,\mathrm{G}$, $P_0=1\,\mathrm{ms}$, $\sin\alpha=1$, $\ell_{\rm Sch}$ from Equation~(\ref{eq:ell_sch}), and the independent parameter $\epsilon_{{\rm gw},0}=10^{-3}$.

Figure~\ref{fig:period_energy} compares four prescriptions. The unit-duty-cycle vacuum curve reproduces the earlier upper bound and is shown only for reference. The two screened curves use $t_{\rm rep}=\ell_{\rm Sch}/c$ and $t_{\rm rep}=R_{\rm LC}/c=P/(2\pi)$, respectively, while the final curve omits the Schwinger term. In the fiducial global-circuit model, the birth values are
$L_{\rm Sp,eff}\simeq8.64\times10^{48}\,\mathrm{erg\,s^{-1}}$,
$L_{\rm gw}\simeq2.44\times10^{48}\,\mathrm{erg\,s^{-1}}$, and
$L_{\rm dp}\simeq2.88\times10^{47}\,\mathrm{erg\,s^{-1}}$. Screening therefore reduces the sustained Schwinger luminosity by about ten orders of magnitude relative to the instantaneous vacuum value. At birth, the screened Schwinger term is comparable to, but larger than, the gravitational-wave term.

The period changes only weakly during the first second in the fiducial screened model:
$P(10^{-2}\,\mathrm{s})\simeq1.000002\,\mathrm{ms}$ and
$P(1\,\mathrm{s})\simeq1.00019\,\mathrm{ms}$. The subsequent evolution is controlled by the competition between the screened Schwinger term, gravitational waves, and dipole radiation. The crossover $L_{\rm Sp,eff}=L_{\rm gw}$ occurs at $t\simeq5.50\times10^{3}\,$s, and the crossover $L_{\rm gw}=L_{\rm dp}$ occurs at
\begin{align}
L_{\rm gw} = L_{\rm dp} \quad \Rightarrow \quad t \simeq 2.40\times10^{5}\,\mathrm{s}.
\end{align}
For the adopted $t_{\rm rep}=R_{\rm LC}/c$ prescription, $L_{\rm Sp,eff}$ declines rapidly as the spin slows. The model loses half of its initial rotational energy at $t\simeq6.51\times10^{3}\,$s and $90\%$ at $t\simeq3.23\times10^{5}\,$s. By $10^8\,\mathrm{s}$, the period has increased to $32.0\,\mathrm{ms}$.

Figure~\ref{fig:luminosities} shows how this result follows from the duty factor. The local-light-crossing prescription gives $f_{\ell/c}\simeq2.9\times10^{-8}$ at birth and therefore remains an optimistic screened upper limit. The global light-cylinder prescription gives $f_{\rm LC}\simeq2.5\times10^{-10}$ at birth. Both factors rise only after the Schwinger rate has dropped and the discharge is no longer the dominant instantaneous vacuum channel.

Over the integration interval $10^{-8}$--$10^8\,\mathrm{s}$, the emitted energies in the fiducial global-circuit model are
\begin{align}
E_{\rm Sp} &\simeq 1.12\times10^{52}\,\mathrm{erg},\\
E_{\rm gw} &\simeq 1.23\times10^{52}\,\mathrm{erg},\\
E_{\rm dp} &\simeq 6.12\times10^{51}\,\mathrm{erg}.
\end{align}
These integrated energies should be read as screened, parameterized estimates, not as a substitute for a kinetic calculation. Including screening removes the unphysical sustained $10^{58}\,\mathrm{erg\,s^{-1}}$ Schwinger luminosity while leaving a birth-time Schwinger contribution of order the other early loss channels.

\section{Implications, caveats, and extensions}
\label{sec:implications}

The main astrophysical implication is that a newborn magnetar may briefly enter a dense pair phase through direct vacuum breakdown rather than only through surface charge extraction, gap formation, and cascade multiplication. Schwinger discharge can therefore provide a direct way to populate the near-surface magnetosphere with leptons. The screening estimate in Section~\ref{sec:screening} shows that this phase cannot be treated as steady vacuum evolution. Its observable impact depends on how the global circuit reopens gaps, closes currents, and converts the initially created pair plasma into an outflow relevant for GRBs and related relativistic transients \citep{2007ApJ...659..561M,2011MNRAS.413.2031M,2015PhR...561....1K}.

A possible high-energy implication follows from the fact that, in the present unscreened benchmark, the pair-production capacity exceeds the classical Goldreich--Julian value by many orders of magnitude and develops in a much stronger electromagnetic environment. This may favor the production of high-energy particles during the first discharge or during later gap-reopening episodes. The polar direction is especially interesting, because particles accelerated there may have a better chance to escape.  A quantitative treatment of particle escape, cascade development, screening,  and the resulting high-energy emission is deferred to a subsequent paper.

Several caveats should be kept in mind. First, the aligned vacuum dipole field used here is a benchmark that maximizes the unscreened parallel electric field before plasma screening develops. Once discharge begins, the magnetosphere will not remain vacuum-like. Time-dependent gaps, pair-loaded currents, and quasi-force-free regions should regulate the actual field through screening and current closure \citep{2007ApJ...657..967B,2013MNRAS.429...20T}. Our calculated Schwinger rate should therefore be regarded as an upper estimate during the initial unscreened phase. The same screening that suppresses $E_\parallel$ may help support the current system responsible for the global dipole torque, so the actual torque is better viewed as a pair-loaded electromagnetic torque with an order-unity geometry-dependent coefficient rather than as a dipole channel switched off by Schwinger discharge.

Second, our adopted Schwinger luminosity $L_{\rm Sp}=2m_{\rm e}c^2\dot N_{\pm}$ is conservative for a fixed pair rate, because it ignores the kinetic energy and radiation emitted after pair creation. The true electromagnetic drain associated with a given unscreened pair rate could therefore be larger than shown here. On the other hand, screening can reduce the rate itself or make the discharge intermittent. These two effects bracket the problem: the unscreened aligned vacuum field gives an upper envelope for the instantaneous pair rate, while the rest mass luminosity gives a lower envelope for the energy cost per created pair. A full kinetic treatment that couples pair creation, field screening, and the global MHD outflow is the natural next step.

Third, to heuristically elaborate on all the effects of spin-down, the hybrid nature of the spin-down model is used. The Schwinger rate is mainly and explicitly analyzed for the aligned vacuum configuration but only numerically computed for misaligned vacuum configurations, whose analytical form is beyond the current work. To compare the energy loss due to the Schwinger process with magnetic dipole radiation, we adopt the standard oblique-rotator torque and use the misaligned angle as a parameter, in which the magnetic-dipole term disappears in the aligned case. 

Fourth, to further compare the emission from Schwinger process and electromagnetic radiation
with that of gravitational waves, we adopt the simplest ellipsoidal model often used for continuous emission from isolated objects, whose ellipticity parameter $\epsilon_{\rm gw}$ is assumed constant. If the deformation is magnetic and decreases with spin, cooling, or field rearrangement, then the gravitational wave luminosity will fall faster than shown here. In fact, the ellipsoidal figure of equilibrium that is energetically balanced between rotation and magnetic field, as the star spins down, will have an increasing magnetic deformation relative to a decreasing rotational deformation, and the boundary in Figure~\ref{fig:birth_maps} can be recomputed within the same framework.

In addition, observed recycled systems demonstrate that neutron stars can operate at millisecond periods and switch between accretion-powered and rotation-powered states \citep{2013Natur.501..517P}. Whether Schwinger discharge can occur in such older binary systems depends on the available magnetic field and on the local unscreened voltage; the present newborn-magnetar calculation does not establish that regime.

\section{Conclusions}
\label{sec:conclusions}
Within a framework based on an unscreened vacuum dipole field, we derived an integral expression and Pad\'{e} approximants for the total Schwinger pair creation rate, together with analytical results in the onset and supercritical limits. The calculation is an upper-bound estimate of the initial pair-creation capacity before plasma screening. For the fiducial model with $B_{\rm p}=10^{14}\,$G and $P_0=1\,$ms, the global near-surface vacuum production rate exceeds the Goldreich--Julian open-field flux by $\sim 9.0\times10^{25}$. Across the $(B_{\rm p},P)$ plane, the equality contour $\dot N_{\pm}=\dot N_{\rm GJ}$ appears around $\chi\sim0.05$, well within the onset regime.

The total pair creation rate exhibits two asymptotic behaviors: an exponentially suppressed onset regime, $\dot N_{\pm}\propto\beta^4\epsilon_{\Omega}^3\exp[-\pi/(\beta\epsilon_{\Omega})]$, and an algebraic supercritical regime, $\dot N_{\pm}\propto\beta^2\epsilon_{\Omega}$. The fiducial model lies close to threshold, at $\chi_0\simeq0.57$, indicating that a large initial pair-production rate does not require a deeply supercritical configuration. The same rate also implies rapid backreaction: the polar Schwinger layer has $\ell_{\rm Sch}\simeq4.13\times10^4\,\mathrm{cm}$, the charge density needed to screen $E_{\parallel}$ is supplied on $t_{\rm fill}\sim6\times10^{-32}\,$s, and the pair-plasma response time is $\omega_{\rm p}^{-1}\sim4\times10^{-14}\,$s. These times are far shorter than the light-crossing and rotation times.

When the screened Schwinger luminosity is included together with magnetic dipole and gravitational-wave losses, the fiducial global-circuit model with $t_{\rm rep}=R_{\rm LC}/c$ gives $L_{\rm Sp,eff}\simeq8.64\times10^{48}\,\mathrm{erg\,s^{-1}}$ at birth, comparable to $L_{\rm gw}\simeq2.44\times10^{48}\,\mathrm{erg\,s^{-1}}$. For the fiducial case with constant $\epsilon_{\rm gw}=10^{-3}$, the integrated energies over $10^{-8}$--$10^8\,\mathrm{s}$ are $E_{\rm Sp}\simeq1.12\times10^{52}$\,erg, $E_{\rm dp}\simeq6.12\times10^{51}$\,erg, and $E_{\rm gw}\simeq1.23\times10^{52}$\,erg. These values depend on the replenishment prescription, but they no longer assume a unit-duty-cycle vacuum magnetosphere.

We therefore draw the following conclusions.

Within the aligned vacuum dipole picture considered here, Schwinger discharge can become the dominant initial pair source in newborn millisecond magnetars once $\chi=\beta\epsilon_{\rm \Omega}$ approaches or exceeds unity. Sustained rotational-energy loss requires a duty cycle set by screening, gap reopening, and current closure.

The actual discharge is unlikely to remain steady. The fallback accretion since the formation of magnetar, the gap formation, and current closure may drive time-dependent, intermittent, or episodic behavior rather than a continuous vacuum discharge. Determining that regime requires a kinetic treatment coupled to the global magnetospheric evolution. 

Compared with the classical Goldreich--Julian mechanism, the present unscreened Schwinger benchmark produces a vastly larger number of pairs and operates in a much stronger electromagnetic field. This may enhance the production of high-energy particles during the first discharge or later intermittent episodes. In particular, particles accelerated along the polar direction may have a better chance to escape, but a quantitative study of particle escape and the resulting high-energy emission must include screening and is deferred to a subsequent paper.

\section*{Acknowledgments}
The authors would like to thank the anonymous referee for valuable comments.
The work of C.M.K. and S.P.K. was supported in part by the Institute of Basic Science (Grant No. IBSR038-D1) and the work of S.P.K. was also supported in part by ICRANet.

\appendix
\makeatletter
\renewcommand{\theHequation}{appendix.\Alph{section}.\arabic{equation}}
\makeatother

\section{Schwinger Pairs and Quantum Backreaction}
\label{sec:backreaction}

The Dirac equation for a charge $e$ in parallel electric and magnetic fields with the vector potential $\vec{A} = -Ect \vec{e}_z + Bx \vec{e}_y$ has the second-order spin-diagonal component equation
\begin{equation}
\begin{aligned}
\Bigl[ \frac{\partial^2}{\partial t^2}
&+ (k_z + eEt)^2
+ \bigl(m^2 + eB (2n+1 - 2\sigma) + i r eE \bigr)
\Bigr] \\
&\times \varphi_{(r)n}^{(\sigma)}(t) = 0.
\end{aligned}
\end{equation}
where $n$ is the Landau level $(n = 0, 1, \cdots )$ and $\sigma = \pm 1/2$ denotes the spin state along the magnetic field whereas $r=\pm 1$ denotes the spin eigenstate of the electric field. In the in-out formalism in which background fields evolve the in-vacuum to the different out-vacuum which is related by the Bogoliubov transformation, the mean number of Schwinger electron-positrons with $m_e$ and $e \, (>0)$ created in a constant electric field $E$ is given by the Bogoliubov coefficient as
\begin{eqnarray}
|\nu ({\bf p}_{\perp})|^2 = e^{- \pi \frac{(m_e c^2)^2 + ({\bf p}_{\perp} c)^2}{e \hbar c E}},    
\end{eqnarray}
where ${\bf p}_{\perp}$ is the transverse momentum perpendicular to
the electric field. In a Lorentz frame in which $E$ is parallel to a constant magnetic field $B$, the mean number is given by
\begin{eqnarray}
|\nu (n, \sigma)|^2 = e^{- \pi \frac{(m_e c^2)^2 + e \hbar c B(2n+1 - 2 \sigma)}{e \hbar c E}}.    
\end{eqnarray}
Note that the mean number depends only on $n$ and $\sigma$ while it is independent of $r$. Applying a Lorentz transformation, summing over all spin states and Landau levels and multiplying the density of states, the total mean number per unit Compton spacetime volume takes the form
\begin{eqnarray}
\sum_{n, \sigma} |\nu (n,\sigma)|^2 = \frac{1}{4 \pi^2} \frac{{\cal E} {\cal B}}{B_Q^2} \coth \Bigl(\pi \frac{\cal B}{{\cal E}} \Bigr) e^{- \pi \frac{B_Q}{\cal E}},    
\end{eqnarray}
where $B_Q$ denotes the Schwinger (critical) field $B_Q = m_e^2 c^3/e \hbar$. Note that $E_Q = B_Q$ in the Gaussian units. We distinguish the mean number of pairs from the vacuum persistence amplitude that prescribes the probability for the in-vacuum to remain in the same vacuum or the vacuum decay
\begin{eqnarray}
- \sum_{n, \sigma} \ln (1 - |\nu (n,\sigma)|^2 ) = \sum_{n,\sigma}\sum_{k = 1}^{\infty} \frac{1}{k} |\nu (n,\sigma)|^{2k}.    
\end{eqnarray}

In the semiclassical approach electromagnetic fields are classical variables while the electrons and positrons obey the Dirac equation with the background fields. Then, the semiclassical Maxwell equation for the polarization current from Schwinger pairs in electric and magnetic fields is given by  (for a review see Refs.~\citep{Kluger:1992md,2010PhR...487....1R})
\begin{eqnarray}
 \partial_{\alpha} F^{\alpha \beta} =\frac{ 4\pi}{c} \langle J^{\beta} \rangle, \quad J^{\beta} = i \frac{e}{2} [\bar{\psi} (E, B), \gamma^{\beta} \psi (E, B)]. 
\end{eqnarray}
In the second quantization, the asymptotic states in a constant electric field parallel to a constant magnetic field yield
\begin{eqnarray}
\langle J^{0} \rangle = e \sum_{n, \sigma} (|\nu_{+} (n,\sigma)|^2 - |\nu_{-} (n,\sigma)|^2) = 0,
\end{eqnarray}
where $|\nu_{\pm} (n,\sigma)|^2$ is the mean number for positrons and electrons, respectively. The equality means that positrons are created by the same number as electrons and the background field keeps charge neutral. The Schwinger pairs with transverse momentum square $ {\bf p}_{\perp}^2$ or $e \hbar c B(2n+1 - 2 \sigma)$ accelerate along the electric field direction and quickly reach the speed of light. Taking asymptotic behavior of $\varphi_{(r)n}^{(\sigma)} (t)$ when $t \gg t_C$, the current density becomes
\begin{eqnarray} \label{eq:J_z}
J_z = -2 ec \frac{1}{4 \pi^2} \frac{E B}{E_Q B_Q} \coth \Bigl(\pi \frac{B}{E} \Bigr) e^{- \pi \frac{E_Q}{E}} \Bigl(\frac{\lambda_C}{c} \Bigr) \bar{t}, 
\end{eqnarray}
where $\bar{t} = t/t_C$ is the dimensionless time in the Compton time. Here we have summed over the Landau levels and spin states.

To find the electromagnetic fields for a given current, one needs to solve the Maxwell equations. Near the pole in the aligned rotator model, the one-directional variation along the $z$ axis ($\partial_x=\partial_y=0$) is a good approximation, with which the Maxwell equations reduce to
\begin{align}
    \partial_z E_z   &= 4\pi \rho ,   &  \partial_t E_z  &=-4\pi J_z, \nonumber \\
     \partial_z B_z  & = 0,    & \partial_t B_z  &=0, \nonumber \\
     \partial_t B_x  &= \partial_z E_y,  & \partial_t E_y  &=c \partial_z B_x - 4\pi J_y, \nonumber \\
     \partial_t B_y   &= -\partial_z E_x,   &\partial_t E_x   &=- c \partial_z B_y - 4\pi J_x. \nonumber \\
  \label{semi-Maxwell}   
\end{align}
Note that $B_z=B_0={\rm constant}$. Focusing on the longitudinal components and assuming a charge neutrality ($\rho=0$), we can solve $\partial_t E_z  =-4\pi J_z$ with (\ref{eq:J_z}) as the source term albeit perturbatively: 
\begin{equation}
{\bar E}_z(t)={\bar E}_0 + \frac{\alpha}{\pi}|{\bar B}_0 ||{\bar E}_0| \coth{ \left( \pi \frac{|{\bar B}_0|}{|{\bar E}_0|}\right)}e^{-\pi/|{\bar E}_0|} {\bar t}^2.
\end{equation}
Then a rough estimation can be made for the time at which the induced field overcomes the initial field ($E_0<0$), which is a mark for a substantial quantum back reaction:
\begin{equation}
    t_{br} =t_C \sqrt{ \frac{\pi}{\alpha} \frac{e^{\pi/|{\bar E}_0|}}{|{\bar B}_0|}\tanh{ \left( \pi \frac{|{\bar B}_0|}{|{\bar E}_0|}\right)}},
\end{equation}
which is $217t_C$ for the fiducial case ($|{\bar B}_0|=2.27$ and $|{\bar E}_0|= 0.570$). In general, the electric field and the current oscillate: roughly speaking, the electric field oscillates like $\cos$ while the current does like $\sin$, and the oscillation amplitude of the electric field is comparable to $E_0$~\citep{Kluger:1992md}. 

The oscillation of Schwinger pairs due to a pure electric field $E$ can be described by the quantum Vlasov equation for a single particle distribution $F(p_{\parallel}, p_{\perp}, t)$~\citep{Kluger:1992md}:
\begin{eqnarray}
\frac{dF}{dt} = \frac{\partial F}{\partial t} + eE \frac{\partial F}{\partial p_{\parallel}} = S (p_{\parallel}, p_{\perp}, t),  
\end{eqnarray}
where $p_{\parallel}= k_z +eEt$ and the source term for fermions is
\begin{eqnarray}
S (p_{\parallel}, p_{\perp}, t) = \frac{eE(t) \epsilon_{\perp}}{2 \omega^2(p_{\parallel}, p_{\perp}, t)}
\int_{- \infty}^{t} dt' \frac{eE(t') \epsilon_{\perp}}{\omega^2(p_{\parallel}, p_{\perp}, t')} \nonumber\\
\Bigl(1-2F(p_{\parallel}, p_{\perp}, t') \Bigr) \cos \Bigl(2 \Theta (p_{\parallel}, p_{\perp}, t) - \Theta (p_{\parallel}, p_{\perp}, t') \Bigr),  
\end{eqnarray}
where $\omega = \sqrt{p_{\parallel}^2 + \epsilon_{\perp}^2}$ with $\epsilon_{\perp} = \sqrt{m^2 +p_{\perp}^2}$ and
$\Theta = \int^{t} dt' \omega (p_{\parallel}, p_{\perp}, t')$.
The oscillation period is
\begin{eqnarray}
\omega_{pl}^2 &=&  2 e^2 \int \frac{dp_{\parallel}}{2\pi} \frac{d^2 p_{\perp}}{(2\pi)^2} N(p_{\parallel}, p_{\perp}) \frac{\partial}{\partial p_{\parallel}} \Bigl(\frac{p_{\parallel}}{\omega (p_{\parallel}, p_{\perp})} \Bigr) \nonumber\\
&=& \frac{\alpha}{2 \pi^3} \frac{E}{E_Q} e^{- \pi \frac{E_Q}{E}} \Bigl(\frac{c}{\lambda_C} \Bigr)^2.
\label{plasma_freq1}
\end{eqnarray}
In the case of parallel electric and magnetic fields, the transverse momentum integral is replaced by $(eB)/2(2 \pi) \sum_{n, \sigma}$, which gives the oscillation frequency
\begin{eqnarray}
 \omega_{pl}^2= \frac{\alpha}{2 \pi^2} \frac{B}{B_Q} \coth \Bigl(\pi \frac{B}{E} \Bigr) e^{- \pi \frac{E_Q}{E}} \Bigl(\frac{c}{\lambda_C} \Bigr)^2.
 \label{plasma_freq2}
\end{eqnarray}
In the $B=0$ limit the plasma frequency~(\ref{plasma_freq2}) recovers Equation~(\ref{plasma_freq1}) in the pure electric field. The corresponding period is given as $T_{pl}=(2\pi)^{3/2}t_{br}$.
Assuming that Equation~(\ref{plasma_freq2}) is a good local approximation, the plasma oscillation period of the fiducial case is $3.42\times10^3t_C$ near the pole but exponentially increases since $\omega_{pl} \propto e^{- (\pi/2 \epsilon_{\Omega}) (B_Q/B_p) (r/R)^4}$. Thus, the plasma oscillation period is longer by many orders either away from the neutron star surface or for weak magnetic fields or longer rotation period than that of the fiducial model at the pole. 

The electron-positron pairs can be annihilated into photons ($e+e^+ \rightarrow \gamma + \gamma$), leading to a limited mean free path $l$. The total scattering cross section is given as
\begin{equation}
\sigma =\lambda_C^2 \frac{ \pi \alpha^2}{\gamma^2} \ln \left(\sqrt{\frac{2}{\rm e}}\gamma \right)
\end{equation}
for relativistic pairs in the center of momentum frame (${\rm e} \simeq 2.72$)  \cite{Greiner2009quantum}. Using the average density defined as ${\dot n}_\pm l/(2c)$ and the cross section, we can estimate the mean free path:
\begin{equation}
    l=\sqrt{\frac{2c}{{\dot n}_\pm \sigma}}= \lambda_C \sqrt{ \frac{8\pi}{\alpha^2} \frac{ e^{\pi/|{\bar E}}}{{|\bar E|}{|\bar B|}}\tanh\left(\pi \frac{{|\bar B|}}{{|\bar E|}} \right) \frac{\gamma ^2}{\ln\left(\sqrt{\frac{2}{\rm e}}\gamma \right)}}.
\end{equation}
For the fiducial case, the mean free path is extremely short near the pole: $0.013~{\rm cm}$ for 1-GeV pairs and $9.24~{\rm cm}$ for 1-TeV pairs. This short mean free path implies that the Schwinger pair production can triggers QED cascades, a plethora of QED cascade can occur~\citep{berestetskii1982quantum,2010MNRAS.406.1379M}. All these processes in strong field  QED plasma of pairs together with gammas will be addressed in a future work.

\section{Asymptotic expansions in the onset and supercritical regimes}
\label{sec:asymptotics}

The asymptotic behavior of the total pair creation rate is governed by the single combination
$\chi=\beta\epsilon_{\rm \Omega}$. We therefore consider separately the limits $\chi\ll1$ and $\chi\gg1$.

\subsection{Onset regime: \texorpdfstring{$\beta\epsilon_{\rm \Omega} \ll 1$}{beta epsrot << 1}}

For $\chi\ll1$, the argument $\zeta(y)$ is large throughout the integration domain, and the incomplete gamma function admits the asymptotic form
\begin{equation}
\Gamma(-1,z) = \frac{e^{-z}}{z^2}
\left[1+\mathcal{O}\!\left(z^{-1}\right)\right],
\qquad z\gg1.
\label{eq:Gamma_large_z}
\end{equation}
The integral is controlled by the minimum of the exponent. Since
$\sqrt{1+3y^2}/y^3$ attains its minimum at $y=1$, the onset of pair creation is localized near the magnetic pole.

To evaluate the integral, we expand about $y=1$ by writing $y=1-u$ with $u\ll1$. Near the pole,
\begin{equation}
\begin{aligned}
\sqrt{1+3y^2} &= 2 - \frac{3}{2}u + \mathcal{O}(u^2), \\
\frac{\sqrt{1+3y^2}}{y^3}
&= 2 + \frac{9}{2}u + \mathcal{O}(u^2),
\end{aligned}
\end{equation}
so that
\begin{equation}
\zeta(y) =
\frac{\pi}{\chi} + \frac{9\pi}{4\chi}u
+ \mathcal{O}\!\left(\frac{u^2}{\chi}\right).
\end{equation}
Substituting Equation~(\ref{eq:Gamma_large_z}) into Equation~(\ref{eq:Ndot_exact}) and retaining the leading contribution gives
\begin{equation}
\sqrt{1+3y^2}\,\Gamma[-1,\zeta(y)]
\simeq
2\,\frac{\chi^2}{\pi^2}
\exp\!\left(-\frac{\pi}{\chi}\right)
\exp\!\left(-\frac{9\pi}{4\chi}u\right).
\end{equation}
Performing the remaining integral over $u$ gives
\begin{equation}
\int_0^1 dy\, \sqrt{1+3y^2}\,\Gamma[-1,\zeta(y)]
\simeq
\frac{8\chi^3}{9\pi^3}\,
\exp\!\left(-\frac{\pi}{\chi}\right).
\end{equation}
Therefore,
\begin{equation}
\dot N_{\pm,\rm onset}
=
\frac{cR^3}{9\pi^4\lambda_{\rm C}^4}\,
\beta^4 \epsilon_{\rm \Omega}^3
\exp\!\left(-\frac{\pi}{\beta\epsilon_{\rm \Omega}}\right),
\qquad
\beta\epsilon_{\rm \Omega} \ll 1.
\label{eq:Ndot_asym}
\end{equation}

Equation~(\ref{eq:Ndot_asym}) shows that, in the onset regime, the effective Schwinger field remains subcritical over nearly the entire surface. The factor
$\exp[-\pi/(\beta\epsilon_{\rm \Omega})]$ then strongly suppresses pair creation and confines the emission to the vicinity of the surface pole, where the local field invariant is maximal.

The emitting region is also geometrically restricted. The angular dependence of the exponent implies that the dominant contribution comes from
$u\sim 4x/(9\pi)$. Since $u=1-y\simeq \theta^2/2$ for $\theta\ll1$, the effective polar-cap half-angle scales as
\begin{equation}
\theta_{\rm eff}
\sim
\left(\frac{8x}{9\pi}\right)^{1/2}.
\label{eq:theta_eff}
\end{equation}
Likewise, because the dipolar field decreases outward as $(R/r)^4$, the emitting region is confined to a thin radial shell above the surface. Writing $r=R(1+\xi)$ with $\xi\ll1$, one finds $(R/r)^4 \simeq 1-4\xi$, so the radial thickness of the active layer is of order
\begin{equation}
\Delta r \sim \frac{Rx}{4\pi}.
\label{eq:dr_eff}
\end{equation}
Hence the onset regime is suppressed by the tunneling factor, the narrow polar angular domain, and the thin emitting layer near the surface. These restrictions account for the exponential smallness of the total rate and for the additional factors $\beta^4\epsilon_{\rm \Omega}^3$ in the prefactor.

\subsection{Supercritical regime: \texorpdfstring{$\beta\epsilon_{\rm \Omega} \gg 1$}{beta epsrot >> 1}}

In the opposite limit $\chi \gg 1$, the argument $\zeta(y)$ becomes small over most of the polar hemisphere, so the incomplete gamma function admits the expansion
\begin{equation}
\Gamma(-1,z)
=
\frac{1}{z}+\ln z+\gamma_{\rm E}-1+\mathcal{O}(z),
\qquad z\ll1,
\label{eq:Gamma_small_z}
\end{equation}
with $\gamma_{\rm E}$ the Euler constant. The only place where the small-$z$ expansion fails is the equatorial boundary layer, because for $y\ll1$ one has
\begin{equation}
\zeta(y)\simeq \frac{\pi}{2x}\,y^{-3}.
\end{equation}
The crossover between the supercritical bulk and the last subcritical strip is therefore fixed by $\zeta(y_c)\sim1$, giving
\begin{equation}
y_c \sim \left(\frac{\pi}{2x}\right)^{1/3}.
\label{eq:yc_super}
\end{equation}
Thus, when $\chi\gg1$, most of the hemisphere contributes without exponential suppression and the residual threshold effect is controlled by an equatorial layer of width $\sim \chi^{-1/3}$.

Substituting the small-$z$ expansion into Equation~(\ref{eq:number_rate}) gives the leading behavior
\begin{equation}
\mathcal J(\chi)
=
\pi\int_0^1dy\,\sqrt{1+3y^2}\,\Gamma[-1,\zeta(y)]
=
\frac{\chi}{2}+\mathcal O(\ln \chi),
\qquad \chi\gg1,
\label{eq:J_super_lead}
\end{equation}
so the corresponding leading supercritical pair-creation rate becomes
\begin{equation}
\dot N_{\pm,\rm super}
=
\frac{cR^3}{16\pi^2\lambda_C^4}\,\beta^2\epsilon_{\Omega}
\left[1+\mathcal O\!\left(\frac{\ln(\beta\epsilon_{\Omega})}{\beta\epsilon_{\Omega}}\right)\right],
\qquad \beta\epsilon_{\Omega}\gg1.
\label{eq:Ndot_super}
\end{equation}
The leading $\chi/2$ term is asymptotically correct. In practice, it is not sufficiently accurate in the moderate supercritical interval $1\lesssim \chi\lesssim10^3$ that matters for the paper. We therefore also use a simple approximation that works for both $\chi<1$ and $\chi>1$,
\begin{equation}
\mathcal J_{\rm simple}(\chi)
\equiv
\frac{\chi}{2}
\exp\!\left(-\frac{\pi}{\chi}\right)
\left[1+0.4\left(\frac{\pi}{2\chi}\right)^{1/3}+1.8\left(\frac{\pi}{2\chi}\right)\right]^{-2},
\qquad \chi\ge1,
\label{eq:J_super_simple}
\end{equation}
which keeps the correct $\chi/2$ limit as $\chi\to\infty$, preserves a smooth turn-on near $\chi\sim1$, and reproduces the reduced integral to within about $8.3\%$ over $1\le \chi\le10^3$. The associated rate is
\begin{equation}
\dot N_{\pm,\rm simple}
=
\frac{cR^3}{8\pi^2\lambda_C^4}\,\beta\,\mathcal J_{\rm simple}(\chi),
\qquad \chi\ge1.
\label{eq:Ndot_super_simple}
\end{equation}
The exponential factor in Equation~(\ref{eq:J_super_simple}) controls the threshold turn-on, while the denominator mimics the finite-width equatorial boundary layer. For all quantitative evolution calculations in this paper we still use the Pad\'{e} expression, because it remains uniformly accurate across the full onset-to-supercritical transition.

At the surface, the effective dimensionless field parameter is
\begin{equation}
\bar{\mathcal E}_*(y)
=
\chi\,\frac{y^3}{\sqrt{1+3y^2}}.
\label{eq:Ebar_surface}
\end{equation}
For $\chi\gg1$, this quantity exceeds unity over most of the polar hemisphere, so the tunneling barrier is effectively absent there. Only near the equator does the factor $y^3$ drive $\bar{\mathcal E}_*$ below unity, making the equator the last region to remain subcritical.

The radial structure also changes. At fixed $y$,
\begin{equation}
\bar{\mathcal E}(r,y)
=
\chi\,\frac{y^3}{\sqrt{1+3y^2}}
\left(\frac{R}{r}\right)^4.
\label{eq:Ebar_radial}
\end{equation}
Unlike the onset regime, where emission is restricted to an exponentially thin shell near the stellar surface, the condition $\bar{\mathcal E}(r,y)\gtrsim1$ can be satisfied over a much broader radial interval. The active volume is therefore no longer limited to a thin polar cap layer, but occupies an order unity angular domain extending through a substantial fraction of the near surface magnetosphere.

Once the exponential suppression is removed over most of the emitting region, the total rate is no longer controlled by tunneling. It is instead set by the algebraic prefactor in the local QED rate. Since the angular integral contributes only the geometric factor $\int_0^1 y^3\,dy = 1/4$, the total rate scales as
\begin{equation}
\dot N_{\pm,\rm super} \propto \beta^2\epsilon_{\rm \Omega},
\qquad
\beta\epsilon_{\rm \Omega} \gg 1,
\end{equation}
instead of exponentially.

Equations~(\ref{eq:Ndot_asym}) and (\ref{eq:Ndot_super}) describe the leading structure of the two limits. For $\beta\epsilon_{\Omega}\ll1$, pair creation is a rare tunneling process confined to a thin polar-cap layer and the rate is exponentially suppressed. For $\beta\epsilon_{\Omega}\gg1$, most of the polar hemisphere becomes supercritical and the rate crosses over to the algebraic scaling of Equation~(\ref{eq:Ndot_super}). Quantitatively, however, the simple $\chi>1$ form in Equation~(\ref{eq:Ndot_super_simple}) is mainly a compact interpolation, whereas the corrected Pad\'{e} approximation remains the best choice across the full transition region.

\begin{figure*}[t]
\centering
\includegraphics[width=\textwidth]{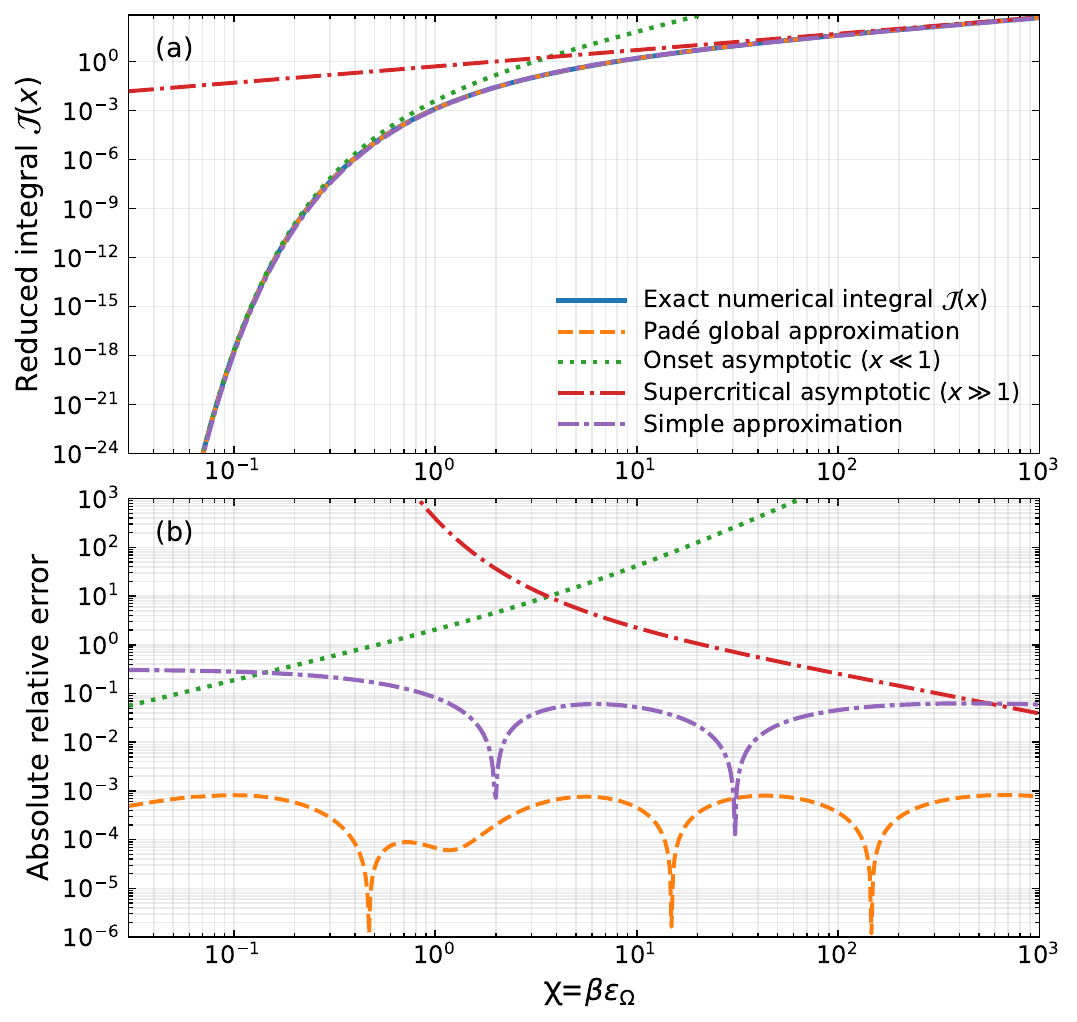}
\caption{Comparison of the numerical \DIFaddbeginFL \DIFaddFL{unscreened }\DIFaddendFL Schwinger rate of Equation~(\ref{eq:number_rate}), the corrected Pad\'{e} rate of Equation~(\ref{eq:sp-rate-approximation}), and the three analytic limits in Equations~(\ref{eq:Ndot_asym}), (\ref{eq:Ndot_super}), and (\ref{eq:Ndot_super_simple}), displayed through their equivalent reduced integrals $\mathcal J(\chi)$. Top: the integral from Equation~(\ref{eq:number_rate}); the Pad\'{e} approximation from Equation~(\ref{eq:sp-rate-approximation}); the green onset asymptotic branch corresponding to Equation~(\ref{eq:Ndot_asym}); the red supercritical asymptotic term $\mathcal J(\chi)\simeq \chi/2$ corresponding to Equation~(\ref{eq:Ndot_super}); and the purple simple $\chi>1$ supercritical fit corresponding to Equation~(\ref{eq:Ndot_super_simple}) or, equivalently, Equation~(\ref{eq:J_super_simple}). All three analytic approximations are shown over the full plotted range for visual comparison. Bottom: absolute relative errors with respect to the integral. The numerical evaluation of the integral uses a guarded incomplete-gamma implementation and clipped exponential factors to avoid floating-point underflow or overflow in the extreme tails. The Pad\'{e} approximation remains below $8.2\times10^{-4}$ throughout $0.03\lesssim \chi \lesssim30$.}
\label{fig:pade_accuracy}
\end{figure*}

Figure~\ref{fig:pade_accuracy} shows the reduced integral $\mathcal J(\chi)$ from Equation~(\ref{eq:number_rate}), the corrected Pad\'{e} approximation, the $\chi<1$ onset asymptotic branch, and the simple $\chi>1$ supercritical fit. The lower panel shows their absolute relative errors with respect to the integral. 

\section{Long-term period evolution in years}
\label{app:longterm}

\begin{figure}[!htbp]
\centering
\includegraphics[width=\columnwidth]{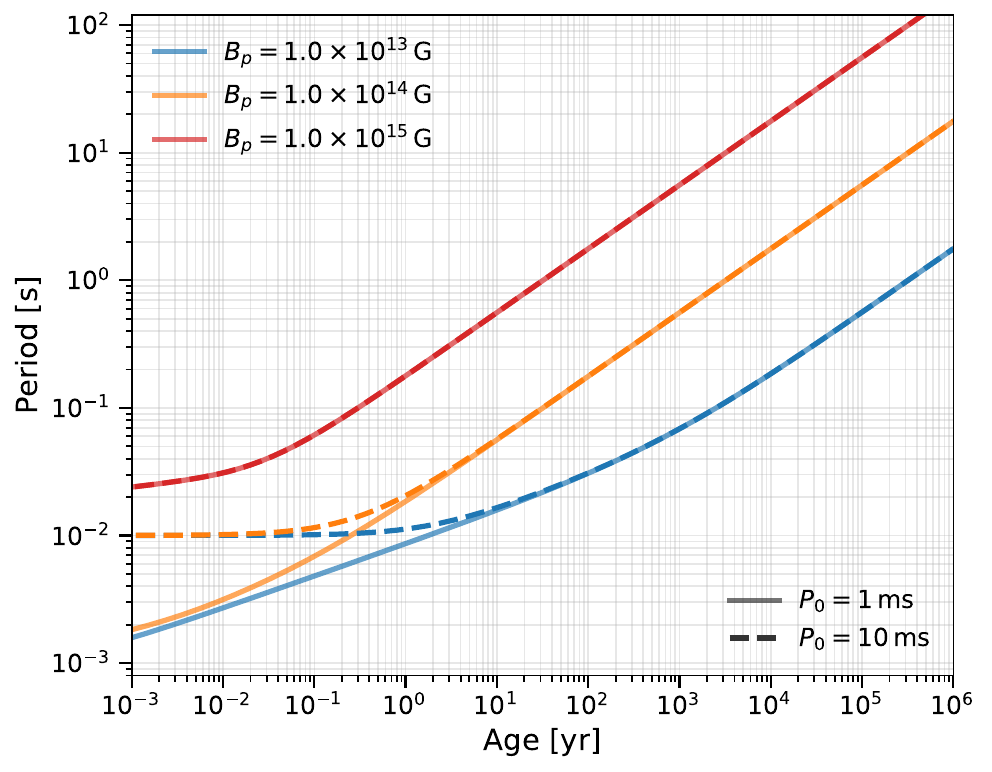}
\caption{Long-term screened spin-period evolution, plotted against age in years and extended to $10^6$ yr, for two birth periods and three surface dipole fields on a single panel. The curves use the global-circuit prescription $t_{\rm rep}=R_{\rm LC}/c$ from Section~\ref{sec:spindown}. The curve color denotes $B_p=10^{13},\,10^{14},\,10^{15}\,$G, while the line style denotes the birth period: solid for $P_0=1\,$ms and dashed for $P_0=10\,$ms. All other stellar parameters are kept fixed at the fiducial values used in Section~\ref{sec:spindown}: $R=12\,$km, $I=1.5\times10^{45}\,\mathrm{g\,cm^2}$, $\sin\alpha=1$, $\ell_{\rm Sch}=R/(7+4\pi/\chi)$, and $\epsilon_{\rm gw}=10^{-3}$. The fiducial model corresponds to the solid orange track with $P_0=1\,$ms and $B_{\rm p}=10^{14}\,$G.}
\label{fig:spin_period_years}
\end{figure}

This long-term evolution illustrates one consequence of the screened global-circuit prescription. It should not be used as a direct prediction for present-day spin periods unless a self-consistent model confirms the assumed gap-reopening cycle and magnetic-field history. The appendix figure is therefore a diagnostic of the spin drain allowed by this parameterized screening model, not a substitute for kinetic magnetospheric evolution.

\clearpage
\bibliographystyle{elsarticle-num}
\bibliography{references}

\end{document}